# Conditional diffusion denoising probabilistic model for super-resolution of atmospheric boundary layer large eddy simulation

Omar Sallam[*,a], Mirjam Fürth[a]

[a]Department of Ocean Engineering, Texas A & M University, College Station, TX, 77843, USA

**ABSTRACT**

Climate change presents a global imperative to accelerate the transition to renewable energy sources. Wind energy, in particular, offers a compelling solution due to its scalability, consistency, and low environmental impact. However, accurate prediction of wind loads and power generation remains a central challenge in wind turbine design and operation. These challenges stem largely from uncertainties in wind shear profiles and turbulence stresses experienced during both normal atmospheric boundary layer (ABL) conditions.

Traditionally, reducing these uncertainties relies on high-fidelity Large Eddy Simulations (LES), which, despite their accuracy, remain computationally prohibitive for large-scale or time-sensitive applications. This work addresses this limitation by leveraging generative machine learning, specifically Conditional Denoising Diffusion Probabilistic Models (DDPMs), to reconstruct high-resolution turbulent flow fields from coarse-resolution inputs.

A high-fidelity numerical dataset is generated using a high performance parallel high-order finite-difference solver across a range of geostrophic wind speeds and surface roughness conditions aligned with IEC wind classes, and at multiple grid resolutions. The diffusion model is trained to super-resolve turbulent structures for different scale factors and is evaluated under both interpolation and extrapolation scenarios.

Results demonstrate that the proposed model accurately recovers fine-scale turbulent structures, Reynolds stresses, and statistical characteristics in interpolation tasks, indicating strong physical consistency within the training regime. However, in extrapolation to higher geostrophic wind speeds, the model exhibits increased noise levels and overprediction of turbulent stresses, highlighting limitations in generalization beyond the training domain.

Overall, this study demonstrates the potential of physics-informed generative models for turbulent flow super-resolution. Such approaches can significantly reduce computational cost while maintaining acceptable accuracy, with direct implications for wind energy applications requiring rapid and reliable turbulent inflow characterization.

## 1. Introduction

The accelerating pace of global climate change presents an urgent challenge to modern societies. Rising temperatures, more frequent and intense extreme weather events such as heatwaves, floods, and wildfires, rapid glacier retreat, sea-level rise, shifting precipitation patterns, and increasing ocean acidification collectively underscore the need to transition away from fossil fuels toward clean and renewable energy sources. Fossil fuel combustion remains the primary driver of greenhouse gas emissions, accounting for nearly 74% of total U.S. emissions [1]. Expanding renewable energy deployment is therefore essential to mitigating long-term climatic and environmental impacts.

Among renewable technologies, wind energy stands out due to its scalability, cost competitiveness, and minimal environmental footprint. Projections indicate that deploying 30 gigawatts (GW) of offshore wind capacity by 2030 could supply electricity to roughly 10 million homes and offset 78 million metric tons of $CO_2$ emissions annually, equivalent to removing approximately 17 million vehicles from U.S. roads [2]. The wind sector has also become a major contributor to the national economy, supporting over 300,000 jobs across all 50 states [3]. In eight states, wind power provides more than 25% of electricity generation, and this share continues to grow as additional projects are commissioned [3]. In 2024 alone, wind energy avoided 351 million metric tons of $CO_2$ emissions, demonstrating its crucial role in decarbonizing the national energy portfolio [3]. With its potential to reduce emissions, strengthen energy security, and stimulate economic growth, wind energy remains a cornerstone of the global transition toward sustainable power systems.

[*]Corresponding author
ORCID(s):

Despite its promise, modern wind energy systems face significant technical challenges that affect their efficiency, reliability, and cost. Wind turbines are continuously exposed to complex, turbulent inflow fields that impose unsteady aerodynamic loads on blades, towers, and nacelles. These fluctuating loads lead to material fatigue, reduce component lifespan, and increase operational and maintenance costs. Extreme atmospheric events further amplify these stresses, potentially driving turbines to experience severe structural loading or even catastrophic failure.

The high capital and manufacturing costs associated with turbine deployment add further pressure to optimize structural design processes. Turbine performance, fatigue life, and economic viability are strongly site-dependent, with atmospheric inflow conditions exerting a dominant influence on both energy production and structural loading. Accurate prediction of the atmospheric boundary layer (ABL) inflow is therefore essential for wind resource assessment, turbine control, and structural load estimation. However, predicting wind loads remains exceptionally challenging. Underestimating inflow turbulence may lead to unsafe design loads, while overestimation can result in overly conservative designs that unnecessarily inflate project costs.

A growing body of research emphasizes the sensitivity of wind turbine structural loads to the characteristics of atmospheric inflow. Robertson et al. [4] performed a comprehensive sensitivity analysis on the NREL 5-MW reference turbine [5], a widely adopted benchmark featuring a 126 m rotor and 90 m hub height. Their results showed that turbulence stress intensity in the streamwise direction and vertical shear are the most influential parameters governing both ultimate and fatigue loads. Specifically, 46% of load outliers were attributed to the streamwise turbulence standard deviation, while 26% were linked to vertical shear, highlighting the critical importance of accurately characterizing these inflow properties.

Recent work by Stanislawski et al. [6] further demonstrated that turbulent inflow fluctuations exert a substantially stronger impact on structural loading than the mean shear profile alone. Large-eddy simulation (LES) results revealed that flapwise blade root loads induced by turbulent inflow are three to four times larger than those resulting solely from shear-driven inflow. Their study also showed that variations in the integral length scale, representing the size of energy-containing turbulent eddies, alter aerodynamic thrust, shear loads, and tower base moments in nontrivial ways. Turbulence intensity was found to have an even greater influence on loads than integral length scales, underscoring the need to capture realistic spatial turbulence structures in modern turbine design and load prediction frameworks.

Accurately resolving these inflow parameters is therefore essential for improving turbine design, forecasting power generation, and planning effective maintenance strategies. However, obtaining high-quality inflow characterizations through field experiments is often impractical due to operational, logistical, and financial constraints. High-fidelity numerical simulations such as fine-resolution LES provide detailed insights into turbulent structures, but they remain computationally prohibitive for routine industrial use, particularly for large domains or long-duration simulations.

To address these limitations, recent advances in machine learning, especially in the context of physics-informed and generative modeling, offer promising pathways for achieving this goal. Stochastic and physics-informed neural networks [7, 8, 9] and conditional probabilistic denoising diffusion models (CPDDMs) [10, 11, 12] provide powerful frameworks for learning multiscale flow structures and capturing the stochastic nature of turbulence.

In this study, we investigate the capability of conditional diffusion-based super-resolution models to recover high-fidelity ABL inflow fields, including vertical shear and turbulence stresses, from coarse-resolution LES inputs under neutral atmospheric conditions. Our goal is to develop an efficient, data-driven surrogate capable of preserving essential physical structures of wall-bounded turbulence while significantly reducing computational costs, thereby enabling more accurate and accessible inflow generation for next-generation wind energy applications.

### 1.1. Large-Eddy Simulation for Atmospheric Boundary Layer Flows

Over the past five decades, large-eddy simulation (LES) has become one of the most powerful and widely used numerical approaches for studying turbulent transport in the atmospheric boundary layer (ABL) [13]. Continuous advances in numerical algorithms, subgrid-scale (SGS) modeling, and validation studies have established LES as a trusted framework for investigating multiscale turbulent processes across a broad range of atmospheric conditions.

LES solves the spatially filtered incompressible Navier–Stokes equations, in which the large, energy-containing turbulent motions are directly resolved while the effects of unresolved subgrid scales are modeled. Applying a spatial filter yields the filtered continuity equation

$$\frac{\partial \overline{u_i}}{\partial x_i} = 0, \qquad (1)$$

and the filtered momentum equation

$$\frac{\partial \overline{u_i}}{\partial t} + \overline{u_j}\frac{\partial \overline{u_i}}{\partial x_j} = -\frac{1}{\rho}\frac{\partial \overline{p}}{\partial x_i} + \nu \frac{\partial^2 \overline{u_i}}{\partial x_j \partial x_j} + f_i - \frac{\partial \tau_{ij}}{\partial x_j}. \qquad (2)$$

The SGS stress tensor

$$\tau_{ij} = \overline{u_i u_j} - \overline{u_i}\,\overline{u_j}, \qquad (3)$$

represents the interactions between resolved and unresolved turbulent scales and introduces a closure problem. Because $\tau_{ij}$ depends on correlations of the unresolved flow $(u_i, u_j)$, it cannot be computed directly from the filtered variables. SGS models, such as the classical Smagorinsky model [14], the Lilly model [15], and nonlinear backscatter formulations [16], approximate $\tau_{ij}$ to account for sub-filter dissipation and backscatter effects. Although the SGS model influences the fine-scale end of the spectrum, numerous studies have shown that the dominant ABL statistics are relatively insensitive to specific SGS formulations, reinforcing the robustness of LES as a predictive tool.

LES has been successfully applied across a wide range of atmospheric regimes, including neutral [17, 18], convective [19, 18], and stable boundary layers [18], as well as flows over plant canopies [20] and heterogeneous land surfaces [21]. Its ability to capture coherent structures, shear-driven turbulence, and stratification effects makes it indispensable for advancing fundamental understanding of ABL dynamics and improving wind energy and environmental modeling [19].

Mirocha et al. [19] conducted a comprehensive evaluation of LES performance using three established solvers: the Weather Research and Forecasting (WRF) model [22], the Simulator for Wind Farm Applications (SOWFA) [23], and the HiGrad solver [24]. Their study examined the sensitivity of predicted wind inflow to geostrophic forcing, surface roughness length, numerical discretization, and SGS model choice under both neutral and convective ABL conditions. Across all solvers, LES reproduced observed wind speeds and turbulence intensities with discrepancies smaller than typical measurement uncertainties, demonstrating both accuracy and consistency independent of solver architecture or SGS closure.

Their analysis also showed that numerical discretization and implicit filtering contribute to spectral deviations near the filter cutoff. Increasing grid resolution and raising the order of the advective operator improved representation of the inertial subrange and enhanced variability in resolved small-scale turbulence. These findings highlight the importance of sufficient grid refinement when high-fidelity turbulence features are required, such as for wind turbine load prediction.

Further insight into grid dependence was provided by Sullivan et al. [25], who examined numerical convergence in convective boundary layer simulations. They demonstrated that low-order statistics (means, variances, and fluxes) become effectively grid independent when there is sufficient separation between energy-containing eddies and the filter scale, quantified by the criterion $z_i/(C_s \Delta_f) > 310$, where $z_i$ is boundary layer height, $\Delta_f$ is filter width, and $C_s$ is the Smagorinsky constant. Higher-order moments such as vertical velocity skewness remained sensitive to resolution, emphasizing their role as indicators of simulation fidelity.

Collectively, these studies underscore the reliability of LES for generating physically consistent ABL turbulence, making it the standard for simulating inflow conditions relevant to wind energy applications. Its ability to resolve shear-driven, anisotropic, and multiscale turbulent motions provides a crucial foundation for data-driven super-resolution models that aim to reconstruct high-fidelity flow fields from coarse-resolution inputs.

### 1.2. Scientific Machine Learning for super resolution and wind inflow

In turbulence-resolving simulations such as large-eddy simulation (LES), the subgrid-scale (SGS) stress tensor $\tau_{ij}$ represents the influence of unresolved turbulent motions on the resolved flow field. Accurate characterization of $\tau_{ij}$ is essential for capturing inter-scale energy transfer and maintaining physical fidelity, particularly in coarse-grid LES. Traditional SGS closures, including the Smagorinsky and dynamic models, rely on assumptions of local scale similarity and near-isotropy, which holds true only when the filter scale is sufficiently small relative to the energy-containing eddies. Achieving such resolution, however, is computationally infeasible for many realistic wind energy applications.

Scientific machine learning could be a potential model offer an alternative pathway by constructing data-driven surrogate models capable of either (i) reconstructing $\tau_{ij}$ directly from coarse-resolution flow fields or (ii) inferring the corresponding high-resolution (HR) velocity fields through super-resolution (SR) techniques [26]. These two tasks are intrinsically linked: recovering HR velocity fields provides the gradients needed to compute $\tau_{ij}$, while learning $\tau_{ij}$ enables closure modeling without explicitly resolving the fine-scale motions.

A general super-resolution mapping may be expressed as
$$\mathbf{I}_{HR} = \mathcal{M}_{SR}(\mathbf{I}_{LR}; \mathbf{W}), \qquad (4)$$
where $\mathbf{I}_{LR}$ denotes coarse-resolution inputs, $\mathbf{I}_{HR}$ the reconstructed high-fidelity fields, and $\mathbf{W}$ the learnable parameters. Similarly, a data-driven SGS closure model can be represented as
$$\tau_{ij} = \mathcal{M}_{SGS}(\mathbf{I}_{LR}; \mathbf{W}), \qquad (5)$$
where the anisotropic SGS stresses are inferred from filtered flow variables. Both mappings enrich coarse flow fields with small-scale information and may be interpreted as inverse modeling strategies guided by statistics learned from high-fidelity data.

Machine learning–based super-resolution has emerged as a promising tool for modeling turbulent flows and reconstructing SGS content. Fukami et.al. [27, 28, 29] demonstrated the capability of convolutional neural networks (CNNs) and hybrid downsampled skip-connection multi-scale (DSC/MS) architectures to reconstruct fine-scale structures in two-dimensional cylinder wakes and homogeneous turbulence. Kim et al. [30] used CNNs, conditional GANs (cGANs), and CycleGAN architectures for artificially coarsened LES fields. Notably, for unsupervised LES-to-DNS super-resolution in turbulent channel flows, CycleGAN achieved superior performance due to its cycle-consistency loss, enabling preservation of high-frequency spatial statistics without requiring paired training data.

Despite these advances, most existing studies remain confined to canonical benchmark flows such as homogeneous isotropic turbulence, channel flows, Taylor–Green vortices, or simple bluff-body wakes. While valuable for model verification, such idealized settings do not reflect the complexity of atmospheric boundary layer (ABL) turbulence. Furthermore, many studies rely exclusively on artificially coarsened high-resolution datasets, which introduces a domain-shift problem: models learn to invert artificial downsampling operators rather than reconstructing physically realistic small-scale motions present in true coarse LES fields.

In contrast, machine learning applications in wind energy more commonly target two-dimensional mesoscale to microscale downscaling, owing to the availability of paired low– and high–resolution datasets. Examples include CCSM [31], ERA5 [32], the WIND Toolkit [33], and COSMO [34]. Leveraging such datasets, Stengel et al. [35] employed a GAN to super-resolve 100 km CCSM wind fields to 2 km resolution using WIND Toolkit data as targets. Building on this, Buster et al. [36] developed the open-source Sup3rCC framework, extending the resolution enhancement to 4 km while incorporating additional meteorological variables such as temperature, humidity, and irradiance. Miralles et al. [37] adopted a similar GAN-based strategy but paired ERA5 with the COSMO model, whereas Lin et al. [38] demonstrated the use of UNet architectures [39] for mesoscale wind super-resolution.

More recently, diffusion models are used to super resolve tropical cyclones form Weather Research Forecasting (WRF) model mesoscale simulation to LES [40] . In addition, data-driven modeling has begun to extend into three-dimensional ABL flows. Rybchuk et al. [41] introduced one of the first diffusion-model-based ABL data assimilation framework, treating flow reconstruction from sparse Doppler lidar measurements as an inpainting task. Their latent diffusion model successfully reconstructed realistic turbulent structures even when observational coverage was below 1% of the domain, demonstrating the capacity of generative models to provide physically plausible initial conditions for LES.

Another line of work focuses on learning SGS closures directly. Cheng et al. [42] trained deep neural networks on DNS of convective boundary layers to predict SGS stresses from coarse-grained velocity fields. Their model outperformed traditional SGS closures, particularly in the surface and mixed layers, and maintained strong generalization across different grid resolutions and stability regimes. A posteriori tests showed recovery of the characteristic $-5/3$ inertial-range spectral scaling and acceptable correlation with DNS SGS stresses.

### 1.3. Contribution

Despite recent progress in machine learning–based super-resolution for turbulent flows, applications aimed at reconstructing high-fidelity atmospheric boundary layer (ABL) inflow fields remain limited. The inherent challenges posed by shear-driven, highly anisotropic turbulence, combined with the infeasibility spatiotemporal of paired low- and high-resolution LES datasets, have impeded the development of reliable data-driven inflow generators suitable for wind energy applications with supervised learning framework.

To address these gaps, the present work introduces a conditional generative diffusion modeling framework capable of performing *true* super-resolution of coarse-resolution ABL LES fields. The proposed approach reconstructs physically realistic velocity fluctuations, vertical shear profiles, and turbulence stress from low-resolution inputs. Moreover, the model is conditioned on key scalar atmospheric parameters—such as geostrophic wind speed and surface roughness length—providing explicit physical control over the generated inflow fields. The overall concept is illustrated in Figure 1.

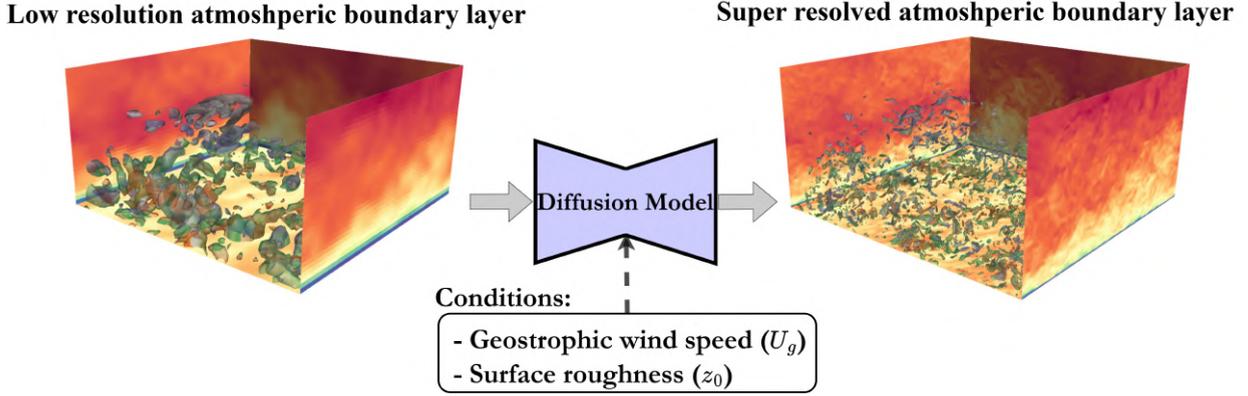

**Figure 1:** Conditional Convolutional UNet architecture for the tropical cyclone mesoscale to LES DDPM super-resolution problem.

## 1.4. Paper Organization

The remainder of this paper is organized as follows. Section 2 describes the numerical configuration for the neutral atmospheric boundary layer, including the generation of LES datasets at multiple resolutions, validation of the numerical setup, and the train–test data partitioning strategy. Section 3 introduces the conditional diffusion model architecture developed for the super-resolution task and details the associated training methodology. Section 4 presents the super-resolution results for both interpolation and extrapolation scenarios, demonstrating the model's capability to reconstruct high-fidelity ABL inflow fields from coarse-resolution inputs. Finally, Section 5 summarizes the main findings and outlines directions for future work.

## 2. LES of ABL and dataset organization

This section presents the model setup and numerical discretization used to simulate the neutral atmospheric boundary layer (ABL) using Large Eddy Simulation (LES). The simulations are performed using Xcompact3D, a high-performance, parallel, high-order finite difference solver [43, 44]. A classical Smagorinsky subgrid-scale (SGS) model is employed to model the unresolved scales of turbulence [14]. The governing equations are the filtered incompressible Navier–Stokes equations, consisting of the filtered continuity and momentum equations:

$$\frac{\partial \overline{u_i}}{\partial x_i} = 0, \tag{6}$$

$$\frac{\partial \overline{u_i}}{\partial t} + \overline{u_j}\frac{\partial \overline{u_i}}{\partial x_j} = -\frac{1}{\rho}\frac{\partial \overline{p}}{\partial x_i} + \nu \frac{\partial^2 \overline{u_i}}{\partial x_j \partial x_j} + \epsilon_{ij3} f(\overline{u_i} - U_{g_i}) - \frac{\partial \tau_{ij}}{\partial x_j}, \tag{7}$$

where $\overline{u_i}$ and $\overline{p}$ are the filtered velocity and pressure fields, respectively; $\nu$ is the kinematic viscosity; $\epsilon_{ij3}$ is the Levi-Civita symbol and $\epsilon_{ij3} f$ represents the Coriolis force action; $f$ is the Coriolis parameter; and $U_{g_i}$ is the component of the geostrophic wind. The last term represents the divergence of the subgrid-scale (SGS) stress tensor $\tau_{ij}$. The filtering operation in LES decomposes the velocity field into resolved and subgrid components:

$$u_i = \overline{u_i} + u'_i, \tag{8}$$

where $u_i$ is the instantaneous velocity, $\overline{u_i}$ is the resolved (filtered) velocity, and $u'_i$ is the subgrid-scale unresolved fluctuation. The SGS stress tensor is defined as:

$$\tau_{ij} = \overline{u_i u_j} - \overline{u_i}\,\overline{u_j}, \tag{9}$$

Using the Smagorinsky model, the SGS stresses are modeled as:

$$\tau_{ij} = -2(C_s \Delta)^2 |\overline{S}|\, \overline{S_{ij}} = -2\nu_{sgs}\overline{S_{ij}}, \tag{10}$$

**Table 1**
IEC wind classes and design wind speeds with extreme events [45, 46].

| IEC Wind Class | I | II | III | IV |
|---|---|---|---|---|
| Annual Average Wind Speed | 10 m/s | 8.5 m/s | 7.5 m/s | 6.0 m/s |
| 50-year Return Gust | 70 m/s | 59.5 m/s | 52.5 m/s | 42 m/s |
| 1-year Return Gust | 52.5 m/s | 44.6 m/s | 39.4 m/s | 31.5 m/s |

where $C_s$ is the Smagorinsky coefficient and $\Delta$ is the filter width, and $\nu_{sgs} = (C_s \Delta)^2 |\overline{S}|$ is the eddy viscosity. The resolved strain-rate tensor $\overline{S}_{ij}$ is given by:

$$\overline{S}_{ij} = \frac{1}{2}\left(\frac{\partial \overline{u}_i}{\partial x_j} + \frac{\partial \overline{u}_j}{\partial x_i}\right), \quad (11)$$

and its magnitude is computed as:

$$|\overline{S}| = \left(2\, \overline{S}_{ij}\, \overline{S}_{ij}\right)^{1/2}. \quad (12)$$

The filter width $\Delta$, which determines the cutoff between resolved and unresolved scales, is defined as:

$$\Delta = (\Delta x \cdot \Delta y \cdot \Delta z)^{1/3}, \quad (13)$$

where $\Delta x, \Delta y, \Delta z$ are the grid spacings in the three spatial directions.

## 2.1. Problem setup

The Large Eddy Simulation (LES) cases cover three geostrophic wind speeds, $U_g = 5.85, 7.15,$ and $10.0$ m/s, which are representative of the 4 ranges of the inflow conditions defined by the International Electrotechnical Commission (IEC) wind classes [45, 46], as summarized in Table 1. The chosen wind speeds are also aligned with inflow conditions commonly used in benchmark simulations of utility-scale benchmark reference wind turbines, including the IEA 15-MW turbine [6], the NREL 5-MW turbine [47, 48], and the DTU 10-MW turbine [49], all of these studies implemented similar ranges of wind speeds. In addition to wind speed variation, the simulations incorporate a range of surface roughness lengths, $z_0 = 0.01, 0.05,$ and $0.1\,[m]$, which correspond to typical terrain classifications where wind energy projects are deployed. These values are selected to represent open flat terrain, low vegetation/agricultural land, and moderately rough terrain, respectively. The choice of $z_0$ values is consistent with previous energy-focused ABL LES studies [48, 6]. The wind speeds and the surface roughness are summarized in Table 2.

The computational domain for all simulations is defined with horizontal dimensions of $L_x = L_y = 2200$ m and a vertical extent of $L_z = 1100$ m. The chosen values for the geostrophic wind speed $U_g$, surface roughness length $z_0$, and domain size are consistent with the parameter ranges established in a comprehensive, collaborative study on large-eddy simulation (LES) of the atmospheric boundary layer (ABL). This study was conducted alongside an extensive experimental measurement campaign targeting wind energy applications [19].

Table 3 summarizes the simulation setup for the LES of the atmospheric boundary layer (ABL), including four different grid resolutions. The coarsest grid configuration, Grid 1, uses $32 \times 32 \times 16$ cells, while the finest grid, Grid 4, employs $256 \times 256 \times 128$ cells. These configurations correspond to horizontal and vertical grid spacings ranging from $68.75\,[m]$ down to $8.4\,[m]$. This selection is consistent with the findings of Wurps et al. [18], who concluded that a grid spacing of approximately $10\,[m]$ and $20\,[m]$ is sufficient to achieve convergence in the first- and second-order flow statistics, as well as turbulent stresses, for both neutral and convective ABL regimes, respectively for similar wind speed ranges.

The table also includes the ratio $h/(C_s \Delta z)$ for each grid configuration, where $h$ is the ABL height, $C_s$ is the Smagorinsky coefficient, and $\Delta z$ is the vertical grid spacing. This ratio indicates the separation between the large-scale production eddies and the subgrid-scale motions that need to be modeled. According to Sullivan et al. [25], low-order moment statistics such as means, variances, and turbulent fluxes become grid independent when this ratio exceeds 310. As shown in Table 3, Grid 1 and Grid 2 fall below this threshold, while Grid 3 and Grid 4 satisfy the criterion.

The time step $\Delta t$ for each simulation is chosen to satisfy the Courant Friedrichs Lewy (CFL) stability condition CFL $\geq 0.25$, where the Courant number is CFL $= \frac{|\mathbf{u}|_{\max} \Delta t}{\Delta x}$. All simulations are carried out for a fixed physical time of 22 hours. The number of CPU cores used and the corresponding wall-clock simulation time for each case are also reported in Table 3.

|  | $z_0 = 0.01[m]$ | $z_0 = 0.05[m]$ | $z_0 = 0.1[m]$ |
|---|---|---|---|
| $U_g = 10.0\ [m/sec]$ | [II-I] / Smooth | [II-I] / Open | [II-I] / Rough |
| $U_g = 7.15\ [m/sec]$ | [IV-III] / Smooth | [IV-III] / Open | [IV-III] / Rough |
| $U_g = 5.85\ [m/sec]$ | IV / Smooth | IV / Open | IV / Rough |

**Table 2**
Wind speeds and surface roughness length test matrix labeled with the wind class and terrain category for the LES ABL simulation.

|  | Grid 1 | Grid 2 | Grid 3 | Grid 4 |
|---|---|---|---|---|
| **Grid Size** | $32 \times 32 \times 16$ | $64 \times 64 \times 32$ | $128 \times 128 \times 64$ | $256 \times 256 \times 128$ |
| **Number of Cells $\times 10^6$** | $\approx 0.0164$ | $\approx 0.131$ | $\approx 1.05$ | $\approx 8.4$ |
| $\Delta x = \Delta y = \Delta z\ [m]$ | $\approx 68.75$ | $\approx 34.38$ | $\approx 17.18$ | $\approx 8.4$ |
| $h/C_s \Delta z$ | $\approx 106.3$ | $\approx 212.6$ | $\approx 425.2$ | $\approx 850.3$ |
| $\Delta t\ [sec]$ | 0.8 | 0.8 | 0.4 | 0.1 |
| **Simulation Time $T$ [hr]** | 22.22 | 22.22 | 22.22 | 22.22 |
| **CPU Cores** | 32 | 32 | 64 | 384 |
| **Wallclock Time [hr]** | 0.023 | 0.07 | 1 | 40 |

**Table 3**
Computational domain spatial and temporal parameters for the LES ABL simulation at different grid sizes.

## 2.2. Validation and verification

The numerical simulation setup is verified and validated using both high-fidelity numerical simulations and field measurements, with all comparisons performed at the finest grid resolution (Grid 4). For a geostrophic wind speed of $U_g = 7.15$ m s$^{-1}$ and a surface roughness length of $z_0 = 0.1$ m, the LES results are verified against the reference numerical simulations reported by Mirocha [19]. Figure 2a compares the predicted vertical turbulent kinematic stress, $u_*^2$, and turbulent kinetic energy (TKE) profiles, demonstrating good agreement with the benchmark results.

For the higher wind speed case, the model is further validated using field measurements presented by Mirocha [19], collected at the Scaled Wind Farm Technology (SWiFT) facility at Sandia National Laboratories [50]. The measurements were obtained using a 200 m tall meteorological tower, providing detailed vertical profiles of atmospheric flow statistics.

As shown in Figure 2b, the LES predictions exhibit strong agreement with the measurements at higher altitudes. Near the surface, the simulations tend to overestimate the measured values, a discrepancy that has also been reported in previous studies. It is important to note that the uncertainty associated with the field measurements, particularly for the velocity components, is relatively high, as discussed in [19].

## 2.3. Parameter variation

In this section, the influence of key numerical simulation parameters on the LES of the atmospheric boundary layer (ABL) is investigated. Specifically, the effects of grid resolution (Section 2.3.1), geostrophic wind speed $U_g$ (Section 2.3.2), and surface roughness length scale $z_0$ (Section 2.3.3) are examined. The analysis presents how variations of these parameters impact the mean wind shear profiles, turbulent stresses, turbulent kinetic energy (TKE), vertical turbulent stresses, and the ratio of the modeled Smagorinsky subgrid-scale (SGS) energy to the total TKE. The variation of $U_g$ and $z_0$ values follows the specifications listed in Table 2, while the different grid resolutions are defined in Table 3.

### 2.3.1. Grid resolution

The influence of grid resolution on large-eddy simulation (LES) of the atmospheric boundary layer (ABL) is examined under multiple geostrophic wind speeds and surface roughness length scales. Four progressively refined grids are considered, denoted as Grid 1 through Grid 4, where Grid 1 represents the coarsest resolution ($32 \times 32 \times 16$) and Grid 4 the finest resolution ($256 \times 256 \times 128$). The corresponding domain configurations are summarized in Table 3. Grid resolution effects are evaluated across three geostrophic wind speeds and three surface roughness lengths, as detailed in Table 2.

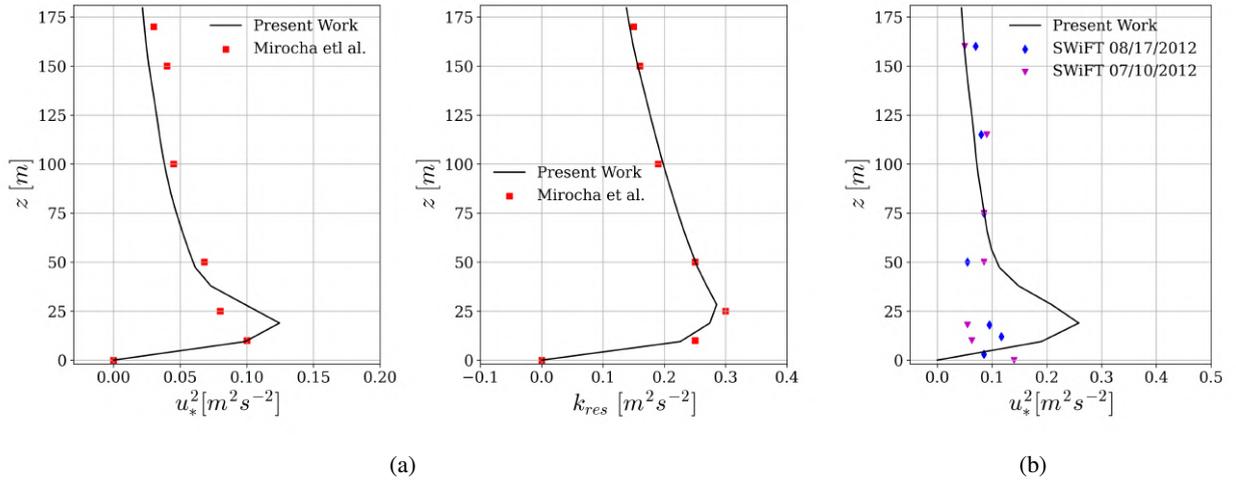

**Figure 2:** Validation and verification of the numerical simulation with previously reported numerical and field experiments.

Figure 3a–3c compare the vertical profiles of the mean streamwise velocity and turbulent stresses obtained using the four grid resolutions for geostrophic wind speeds of $U_g$ = 5.85, 7.15, and 10.0 m s$^{-1}$. The results demonstrate sensitivity of the mean velocity profiles to grid resolution, particularly within the lower portion of the boundary layer up to approximately 200–250 m. In this near-surface region, pronounced discrepancies are observed among the different grids, whereas at higher altitudes the profiles progressively converge and become largely insensitive to resolution.

Across all geostrophic wind speeds considered, coarser grids systematically underestimate the mean wind speed in the lower boundary layer relative to finer grids. This underprediction diminishes with increasing grid refinement, and the differences between successive grids decrease accordingly. In particular, the deviation between Grid 3 and Grid 4 is significantly smaller than that observed between Grid 1 and Grid 2 or between Grid 2 and Grid 3, indicating convergence of the mean velocity profiles at higher resolutions.

A similar resolution dependence is observed for the resolved turbulent stresses $\langle u'u' \rangle$, $\langle v'v' \rangle$, and $\langle w'w' \rangle$, as well as for the friction velocity $u_*$ and the resolved turbulent kinetic energy $K_{\text{res}}$. These quantities exhibit smaller differences between Grid 3 and Grid 4 compared to the larger discrepancies observed between the coarser grids, further suggesting that Grid 3 represents a near-converged LES resolution.

To quantitatively assess LES resolution quality, the dimensionless parameter $\gamma$, defined as the ratio of resolved kinetic energy to the total (resolved plus subgrid-scale) kinetic energy, is evaluated:

$$\gamma = \frac{K_{\text{res}}}{K_{\text{res}} + K_{\text{SGS}}}. \tag{14}$$

The modeled subgrid-scale (SGS) kinetic energy is estimated as

$$K_{\text{SGS}} = \frac{2}{3}(C_s \Delta)^2 |\overline{S}|^2, \tag{15}$$

where $\Delta$ is the filter width, $\overline{S}$ is the magnitude of the resolved strain-rate tensor, and $C_s = 0.14$ is the Smagorinsky coefficient.

The parameter $\gamma$ serves as a widely used indicator of LES resolution adequacy [51]. Values of $\gamma$ approaching unity indicate that the majority of the turbulent kinetic energy is explicitly resolved, whereas lower values reflect increased dependence on the SGS model. As discussed in [52, 51], an LES is generally considered adequately resolved when $\gamma \gtrsim 0.8$.

The vertical profiles of $\gamma$ show that Grid 1 consistently produces values below the recommended threshold of 0.8, indicating insufficient resolution to capture the dominant turbulent scales. In contrast, Grid 3 and Grid 4 maintain $\gamma$ values exceeding 0.8 throughout most of the boundary layer, with $\gamma$ approaching unity, confirming that these grids provide sufficient resolution for well-resolved ABL LES. Grid 2 lies near the lower limit of acceptability, while Grid 1 is clearly under-resolved.

These findings are consistent with prior studies on ABL LES resolution requirements. Sullivan *et al.* [25] reported that a ratio of $h/(C_s \Delta z) \gtrsim 310$ is necessary to ensure adequate vertical resolution for ABL simulations, where $h$ is the boundary layer height and $\Delta z$ is the vertical grid spacing. In the present study, this ratio is approximately 106.3 for Grid 1 and 212.6 for Grid 2, both below the recommended threshold. Conversely, Grid 3 and Grid 4 are 425.2 and 850.3 respectively which exceed this criterion, as shown in Table 3

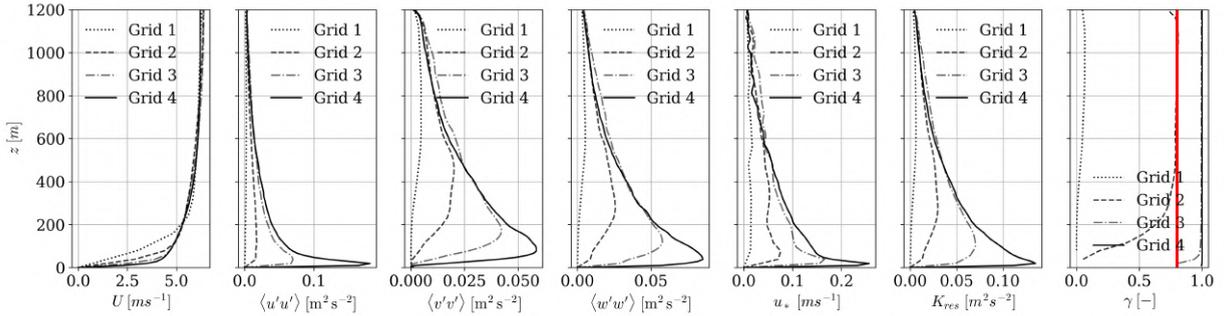

(a) Geostrophic wind speed $U_g = 5.85\ [m/sec]$ and surface roughness $z_0 = 0.01[m]$

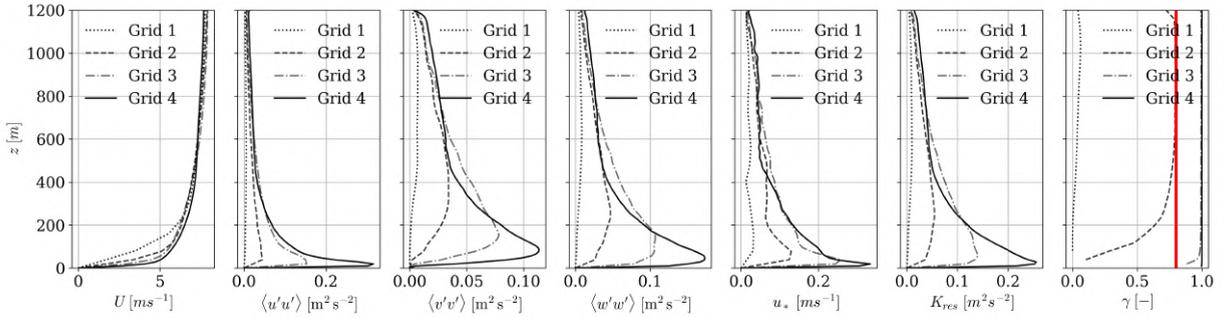

(b) Geostrophic wind speed $U_g = 7.15\ [m/sec]$ and surface roughness $z_0 = 0.05[m]$

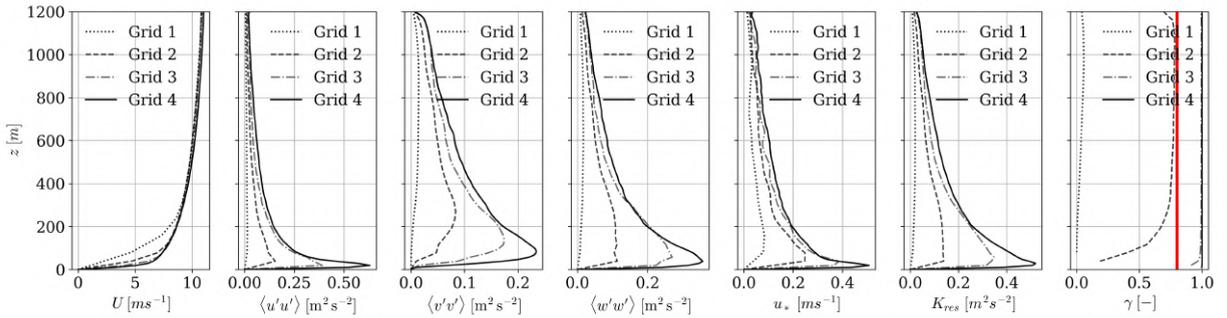

(c) Geostrophic wind speed $U_g = 10.0\ [m/sec]$ and surface roughness $z_0 = 0.1[m]$

**Figure 3**: Parameters variation with grid resolutions at different geostrophic wind speeds.

### 2.3.2. Geostrophic wind speed

The influence of geostrophic wind speed on the mean velocity profiles and turbulent stresses is examined across different surface roughness length scales. Figure 4 presents the variation of the mean streamwise velocity and resolved turbulent stresses for multiple geostrophic wind speeds, where in Figures 4a–4c illustrate the corresponding behavior under different surface roughness conditions.

For all surface roughness values considered, increasing the geostrophic wind speed leads to a systematic increase in the magnitude of the resolved turbulent stresses. This behavior is consistent with the enhanced shear production associated with stronger geostrophic forcing, which intensifies turbulence generation throughout the boundary layer. In addition, higher geostrophic wind speeds are accompanied by a slight upward shift in the height at which the maximum turbulent stress occurs, reflecting a deeper and more energetic boundary layer under stronger synoptic forcing.

Surface roughness further modulates this response. For a given geostrophic wind speed, increasing the surface roughness length results in larger turbulent stress magnitudes, particularly in the near-surface region. This reflects the enhanced momentum extraction and increased shear induced by rougher surfaces. The combined effects of geostrophic wind speed and surface roughness therefore control both the intensity and vertical distribution of turbulence within the ABL, with higher wind speeds and rougher surfaces producing stronger and more vertically extended turbulent structures.

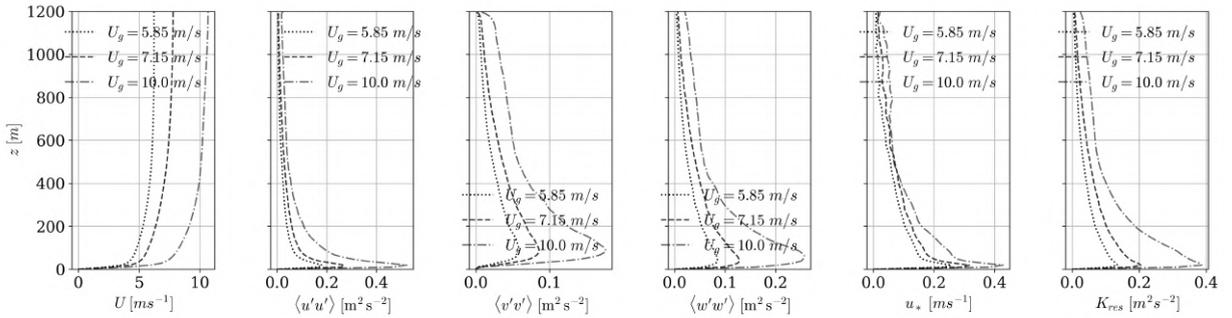

(a) Surface roughness $z_0 = 0.01$ [$m$]

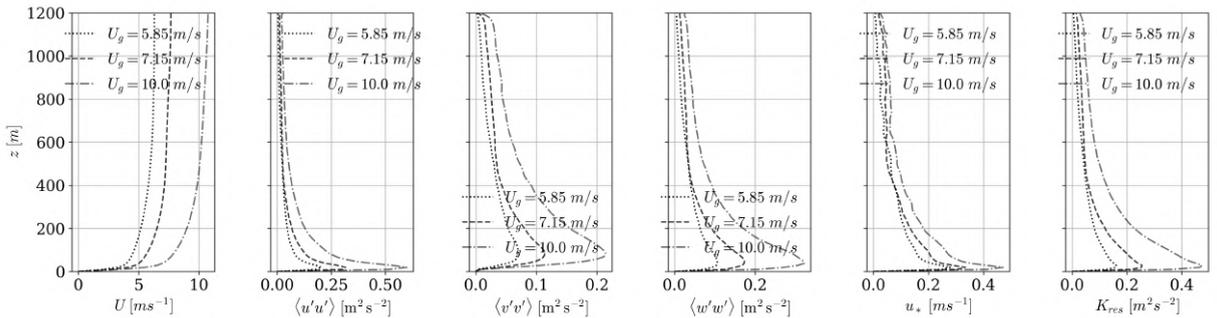

(b) Surface roughness $z_0 = 0.05$ [$m$]

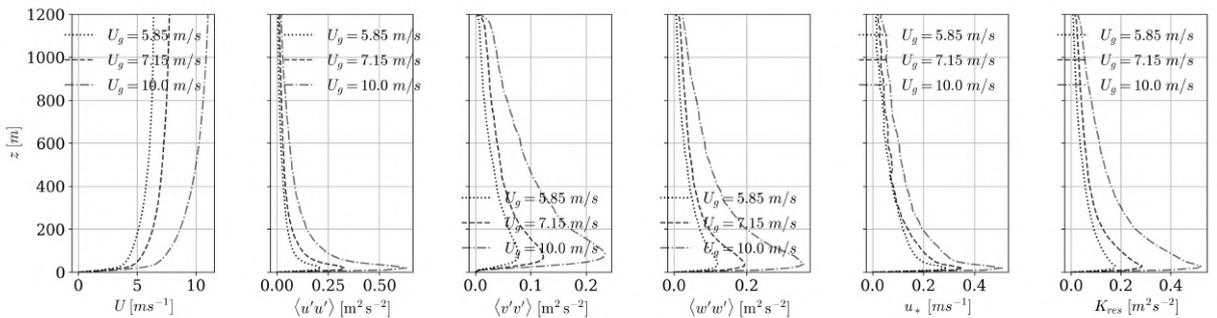

(c) Surface roughness $z_0 = 0.1$ [$m$]

**Figure 4:** Parameters variation with geostrophic wind speeds at different surface roughness values.

| | Test | Test | Train |
|---|---|---|---|
| $U_g = 10.0\ [m/sec]$ | Test | Test | Train |
| $U_g = 7.15\ [m/sec]$ | Test | Test | Test |
| $U_g = 5.85\ [m/sec]$ | Train | Test | Train |
| | $z_0 = 0.01[m]$ | $z_0 = 0.05[m]$ | $z_0 = 0.1[m]$ |

**Table 4**
Training and testing dataset split for the ABL super resolution problem for the interpolation task.

| $U_g = 10.0\ [m/sec]$ | Test | Test | Test |
|---|---|---|---|
| $U_g = 7.15\ [m/sec]$ | Train | Test | Test |
| $U_g = 5.85\ [m/sec]$ | Train | Train | Test |
| | $z_0 = 0.01[m]$ | $z_0 = 0.05[m]$ | $z_0 = 0.1[m]$ |

**Table 5**
Training and testing dataset split for the ABL super resolution problem for the Extrapolation task.

### 2.3.3. Surface roughness coefficient

The influence of surface roughness length on ABL flow characteristics is examined for three geostrophic wind speeds, $U_g = 5.85, 7.15$, and $10.0\ \mathrm{m\,s^{-1}}$. The variations in mean velocity profiles and turbulent stresses with surface roughness are summarized in Figure 5, with detailed results for each geostrophic wind speed shown in Figures 5a, 5b, and 5c, respectively.

At the lower and intermediate geostrophic wind speeds, the mean streamwise velocity profiles exhibit only minor sensitivity to surface roughness over the range of roughness lengths considered. In contrast, at the highest geostrophic wind speed ($U_g = 10.0\ \mathrm{m\,s^{-1}}$), a more pronounced dependence on surface roughness is observed. Specifically, increasing surface roughness leads to a deeper boundary layer, as evidenced by an upward shift in the velocity profile and an increase in the effective boundary layer height (see the left panel of Figure 5c). This behavior reflects enhanced surface drag and momentum extraction under stronger geostrophic forcing.

The turbulent stresses show a clearer and more systematic dependence on surface roughness. For all three geostrophic wind speeds, increasing the surface roughness length results in higher magnitudes of the resolved turbulent stresses and increased resolved turbulent kinetic energy. Additionally, the height at which the turbulent stresses attain their maximum values shifts upward with increasing roughness length. This trend is consistent across all geostrophic wind speeds considered and indicates that rougher surfaces promote stronger shear production and vertically extend the region of active turbulence within the ABL.

### 2.4. Dataset split and preparation

The dataset is prepared and divided into training, validation, and testing splits for both interpolation and extrapolation tasks, as summarized in Tables 4 and 5, respectively. For each case, corresponding to a specific combination of geostrophic wind speed ($U_g$) and surface roughness length scale ($z_0$). A total of 100 temporal frames are extracted from the high-resolution (HR) LES fields. Two-dimensional slices from these HR frames are used as inputs to train the DDPM model.

For the interpolation task, the training dataset consists of two low geostrophic wind speed cases ($U_g = 5.85$ m/s) at $z_0 = 0.01$ m and $z_0 = 0.1$ m, along with one high geostrophic wind speed case ($U_g = 10$ m/s) at $z_0 = 0.1$ m. The remaining case combinations are reserved for testing, see Table 4. For the extrapolation task, training is performed using two low geostrophic wind speed cases ($U_g = 5.85$ m/s) at $z_0 = 0.01$ m and $z_0 = 0.05$ m, and one medium geostrophic wind speed case ($U_g = 7.15$ m/s) at $z_0 = 0.01$ m, with the rest used for testing, see Table 5.

Throughout all tasks, the resolution of the sparse input condition is fixed to a $16 \times 8$ grid (137.5 [m]), independent of the super-resolution downscale factor applied. Training is carried out using a batch size of 20.

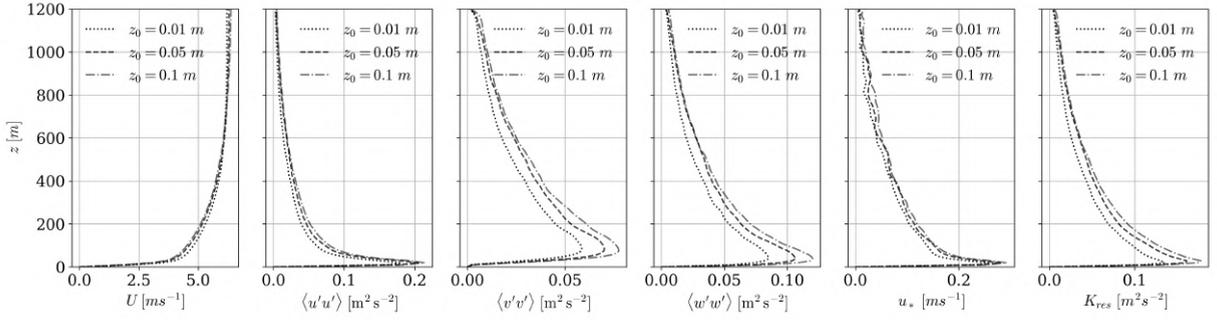
(a) Geostrophic wind speed $U_g = 5.85\ [m/sec]$

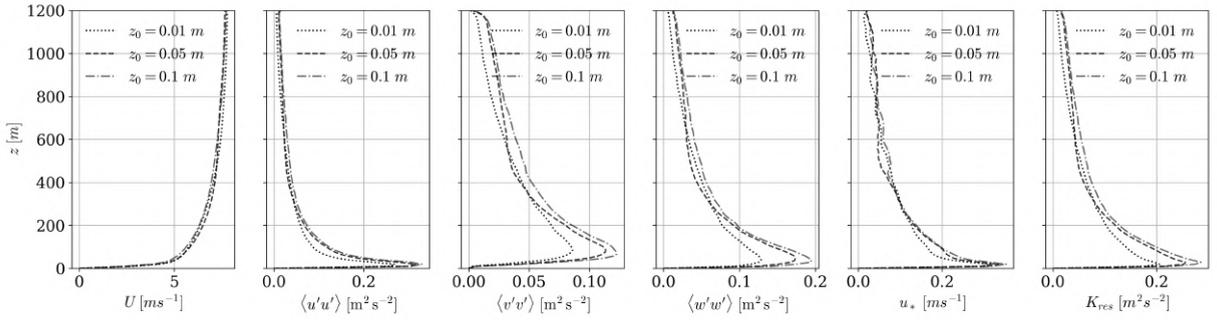
(b) Geostrophic wind speed $U_g = 7.15\ [m/sec]$

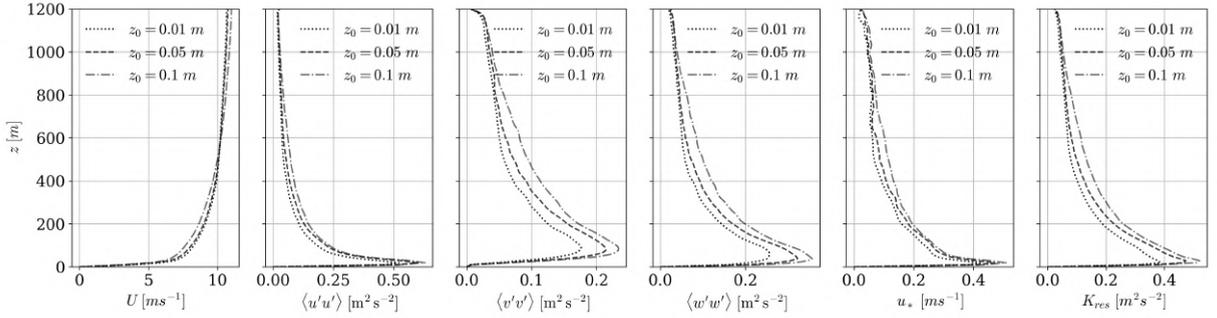
(c) Geostrophic wind speed $U_g = 10.0\ [m/sec]$

**Figure 5:** Parameters variation with surface roughness speeds at different geostrophic wind speeds

## 3. Diffusion-based super-resolution framework for ABL

This section presents a diffusion-based generative modeling framework for super-resolving mesoscale wind fields to LES-resolution using a conditional denoising diffusion probabilistic model (DDPM). We begin with a general overview of the diffusion model fundamentals, followed by a discussion of the proposed architecture, conditioning strategy, and training protocol for super-resolution in the context of tropical cyclones. We end this section with a presentation and discussion of the results of the super-resolution model.

### 3.1. Diffusion models in generative learning

Diffusion models [53] provide a flexible generative learning paradigm that models the data distribution via a two-stage process: (i) a forward diffusion process and (ii) a reverse diffusion process. The forward diffusion process, denoted as $q(\mathbf{x}_t|\mathbf{x}_{t-1})$, gradually adds Gaussian noise to clean data $\mathbf{x}_0 \sim q(\mathbf{x}_0)$, producing a sequence of progressively noisier variables until it reaches $\mathbf{x}_T \sim \mathcal{N}(0, \mathbf{I})$. This process is fixed and non-trainable.

Where at any arbitrary forward timestep $t$, the noisy data sample $\mathbf{x}_t$ is sampled from:

$$\mathbf{x}_t = \sqrt{\bar{\alpha}_t}\mathbf{x}_0 + \sqrt{1-\bar{\alpha}_t}\epsilon, \quad \epsilon \sim \mathcal{N}(0, \mathbf{I}), \tag{16}$$

where $\bar{\alpha}_t = \prod_{s=1}^{t} \alpha_s$ with $\alpha_t = 1 - \beta_t$, and $\{\beta_t \in (0,1)\}_{t=1}^{T}$ is a predefined noise variance schedule. The term $\epsilon$ represents standard Gaussian noise.

The reverse diffusion process, denoted as $p_\theta(\mathbf{x}_{t-1}|\mathbf{x}_t)$, is a learned probabilistic model that aims to invert the noising process and recover the original data distribution from input noisy data. Where time step $t = 0$ is for the clean original data, and time step $t = T$ is for the fully corrupted noisy data.

The reverse process is parameterized by a neural network $\epsilon_\theta(\mathbf{x}_t, t)$, which estimates the noise to be removed at each timestep. Using this prediction, the denoised sample at the previous timestep $\mathbf{x}_{t-1}$ is computed as:

$$\mathbf{x}_{t-1} = \frac{1}{\sqrt{\alpha_t}}\left(\mathbf{x}_t - \frac{\beta_t}{\sqrt{1-\bar{\alpha}_t}}\epsilon_\theta(\mathbf{x}_t, t)\right) + \sqrt{\beta_t}\epsilon. \tag{17}$$

The reverse process $p_\theta(\mathbf{x}_{t-1} | \mathbf{x}_t)$ is modeled as a Gaussian distribution whose mean is learned via the noise prediction network:

$$\boldsymbol{\mu}_\theta(\mathbf{x}_t, t) = \frac{1}{\sqrt{\alpha_t}}\left(\mathbf{x}_t - \frac{\beta_t}{\sqrt{1-\bar{\alpha}_t}}\epsilon_\theta(\mathbf{x}_t, t)\right). \tag{18}$$

The neural network $\epsilon_\theta$ is trained using a simplified loss function derived from the variational bound on the data likelihood, which minimizes the mean squared error between the true noise and the predicted noise:

$$\mathbb{E}_{\mathbf{x}_0, \epsilon}\left[\left\|\epsilon - \epsilon_\theta(\sqrt{\bar{\alpha}_t}\mathbf{x}_0 + \sqrt{1-\bar{\alpha}_t}\epsilon, t)\right\|^2\right]. \tag{19}$$

This formulation has been demonstrated to produce high-fidelity samples and achieve state-of-the-art results on image generation tasks, outperforming GANs in both sample quality and mode coverage [54].

### 3.2. Conditional DDPM for super-resolving ABL

In this work, the conditional denoising diffusion probabilistic model (DDPM) is applied to super-resolve 2D planar sections of coarse-resolution ABL simulations. Each high-resolution training sample undergoes a forward diffusion process that perturbs the original sample $\mathbf{x}_0$ into a prior Gaussian noise distribution $\mathbf{x}_T \sim \mathcal{N}(0, \mathbf{I})$. The model learns to reverse this process via a conditional U-Net that predicts the total noise at each step to reconstruct the original clean data. As shown in Figure 6a, the forward diffusion process gradually corrupts the clean data with Gaussian noise, while the reverse process denoises it until it resembles the original distribution.

The network architecture, illustrated in Figure 6b, is a conditional U-Net [39] with residual convolutional blocks [55] and convolutional attention blocks [56]. The input to the U-Net is the data at an arbitrary noise time step $\mathbf{x}_t$ as 2D tensors, and the network is conditioned on three inputs:

- Sparse low-resolution 2D sections velocity measurements $(u, v, w)$,

- Physical scalar conditions: the geostrophic wind speed and the surface roughness.

- Time step $t$ embeddings computed via sinusoidal Fourier features based MLP,

The time step and scalar parameters are encoded through separate MLPs and injected into every residual block in the encoder, decoder, and bottleneck layers in the U-Net. The sparse velocity field from the mesoscale simulation or the artificially coarsened LES is convolved into the proper feature maps channel depth and injected at the encoder and bottleneck stages of the U-Net. The network starts with a convolution layer mapping the noisy input $\mathbf{x}_t$ to 64 base channels. The U-Net has four encoder-decoder levels with skip connections, halving resolution in each dimension and doubling channels at each stage, reaching a 1024-channel bottleneck. All model inputs are z-score normalized. Training is performed for 50 epochs using the Adam optimizer [57] with a learning rate of $10^{-4}$. Each training batch size consists of 16 samples. The diffusion schedule consists of 500 steps during both training and sampling.

The model is tested in two tasks: interpolation and extrapolation as shown in Section 2.4.

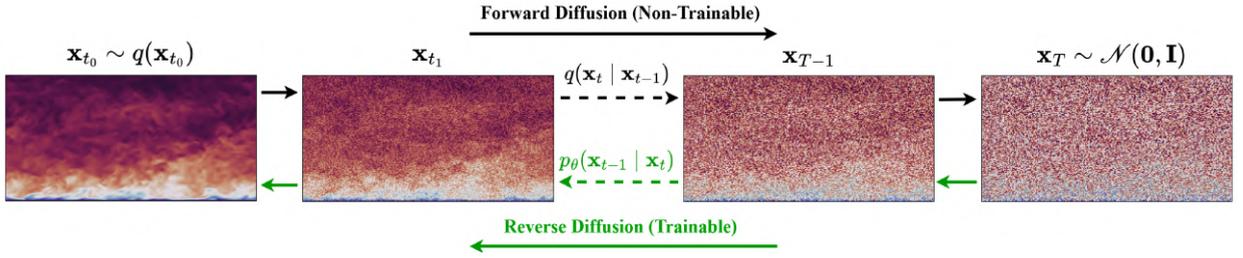

(a) Forward and reverse diffusion processes for the tropical cyclone mesoscale to LES super-resolution problem.

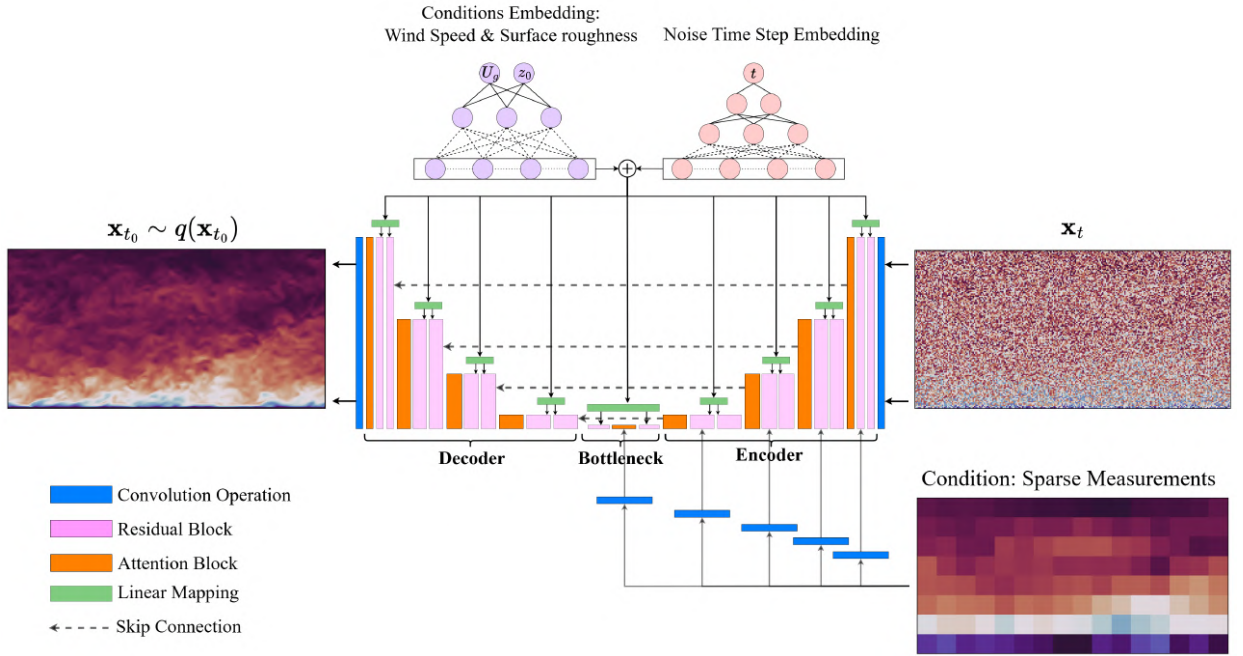

(b) Conditional Convolutional UNet architecture for the tropical cyclone mesoscale to LES DDPM super-resolution problem.

**Figure 6:** Schematic overview of the proposed diffusion-based super-resolution framework for tropical cyclone wind field reconstruction.

## 4. Results

This section presents the super-resolution performance of the proposed framework under both interpolation and extrapolation scenarios. The evaluation is conducted across multiple spatial scale factors (×2, ×4, and ×8), varying geostrophic wind speeds ($U_g$), and different surface roughness coefficients ($z_0$).

### 4.1. Interpolation Task

For the interpolation task, the dataset configuration summarized in Table 4 is considered. Super-resolution results are presented for scale factors ×2, ×4, and ×8 under geostrophic wind speeds of $U_g = 7.15$ m/s and $U_g = 10.0$ m/s. All simulations in this subsection correspond to a surface roughness coefficient of $z_0 = 0.05$ m.

For the scale factor ×2, corresponding to input grid 2 as summarized in Table 3, the velocity contour comparisons are presented in Figures 7a and 8a for geostrophic wind speeds $U_g = 7.15$ m/s and $U_g = 10.0$ m/s, respectively. The figures illustrate the low-resolution input, the high-resolution reference, and the model prediction for all three velocity components.

The predicted fields demonstrate strong visual agreement with the high-resolution snapshots. Large-scale, low-frequency flow structures are consistently inherited from the low-resolution input, while the model reconstructs the missing small-scale turbulent content. The super-resolved fields exhibit coherent fine-scale structures that closely resemble those of the reference LES.

The mean velocity profile reconstruction remains consistent with the low-resolution input grid, as expected for an interpolation setting where large-scale dynamics are preserved. Only minor discrepancies are observed between low- and high-resolution profiles. In contrast, the turbulent stress components show clear improvement in the predicted fields relative to the input grid. Specifically, the reconstructed Reynolds stress terms $u'_i u'_j$, friction velocity $u_*$, and resolved turbulent kinetic energy $k_{\text{res}}$ demonstrate closer agreement with the high-resolution reference, as shown in Figures 7b and 8b.

Similarly, the energy spectrum reconstruction aligns well with the high-resolution spectrum across the resolved wavenumber range at different altitudes (Figures 7c and 7d for geostrophic wind speeds $U_g = 7.15$ m/s and Figures 8c and 8d for geostrophic wind speeds $U_g = 10.0$ m/s ). The predicted spectra recover the inertial subrange more accurately than the low-resolution input, showing the model's ability to restore small-scale turbulent energy while maintaining consistency with the large-scale flow dynamics.

For the interpolation task at scale factor ×4, the model maintains consistent super-resolution performance despite the increased resolution gap. As illustrated in Figures 9 and 10, the predicted fields demonstrate improved reconstruction of velocity contours, turbulent stresses, and energy spectra relative to the low-resolution input for both geostrophic wind speeds ($U_g = 7.15$ m/s and $U_g = 10.0$ m/s).

The large-scale flow structures remain governed by the input grid, while the model restores fine-scale turbulent features and recovers a substantial portion of the missing spectral energy, showing strong agreement with the high-resolution reference solution.

For scale factor ×8, corresponding to input grid 0 as summarized in Table 3, the super-resolution task becomes significantly more challenging due to the severe loss of small-scale information in the low-resolution input. As shown in Figures 11a and 12a, the reconstructed velocity fields remain largely constrained by the coarse input structures, which exhibit poor $\gamma$ values and limited spatial variability. Although the model attempts to restore fine-scale turbulent content, the recovered fields do not fully reproduce the coherent high-resolution structures observed in the reference LES.

The turbulent stress profiles further reflect this degradation. As illustrated in Figures 11b and 12b, the predicted Reynolds stress components exhibit noisy behavior and multiple altitude peaks, which are physically inconsistent with the expected boundary-layer structure. While the mean velocity profile shows improved alignment with the high-resolution solution compared to the low-resolution input, its shape deviates from the physically consistent monotonic structure, indicating instability in the reconstruction at this scale separation.

The energy spectrum at lower altitudes ($z = 100$ m) further illustrates the reconstruction behavior at scale factor ×8. As shown in Figures 11c and 12c for $U_g = 7.15$ m/s and $U_g = 10.0$ m/s, respectively, the predicted spectrum exhibits elevated energy levels in the large-scale (low-wavenumber) region compared to the high-resolution LES. This amplification is attributed to the excessively smooth low-resolution input near the ground, which alters the distribution of large-scale energy.

In contrast, within the small-scale (high-wavenumber) region, the reconstructed spectrum demonstrates improved alignment with the high-resolution reference. Despite the severe resolution gap, the model is able to partially recover high-frequency turbulent energy, indicating that small-scale reconstruction remains feasible even when large-scale consistency is compromised.

Figures 13 and 14 present the isosurfaces of the $Q$-criterion for the low-resolution LES input, the model reconstruction, and the high-resolution reference across different input grid resolutions for geostrophic wind speeds $U_g = 7.15$ m/s and $U_g = 10.0$ m/s. The $Q$-criterion provides a qualitative assessment of coherent vortical structures and small-scale turbulent organization within the boundary layer.

At lower input resolutions, the reconstructed isosurfaces exhibit reduced structural density and weaker vortex coherence, reflecting the limited turbulent information available in the coarse input field. As the input grid resolution increases (corresponding to smaller scale factors), the reconstructed $Q$-criterion structures become progressively denser and more spatially organized, approaching the topology and distribution observed in the high-resolution LES. This trend confirms that the model's ability to recover coherent turbulent structures improves systematically with increasing input resolution.

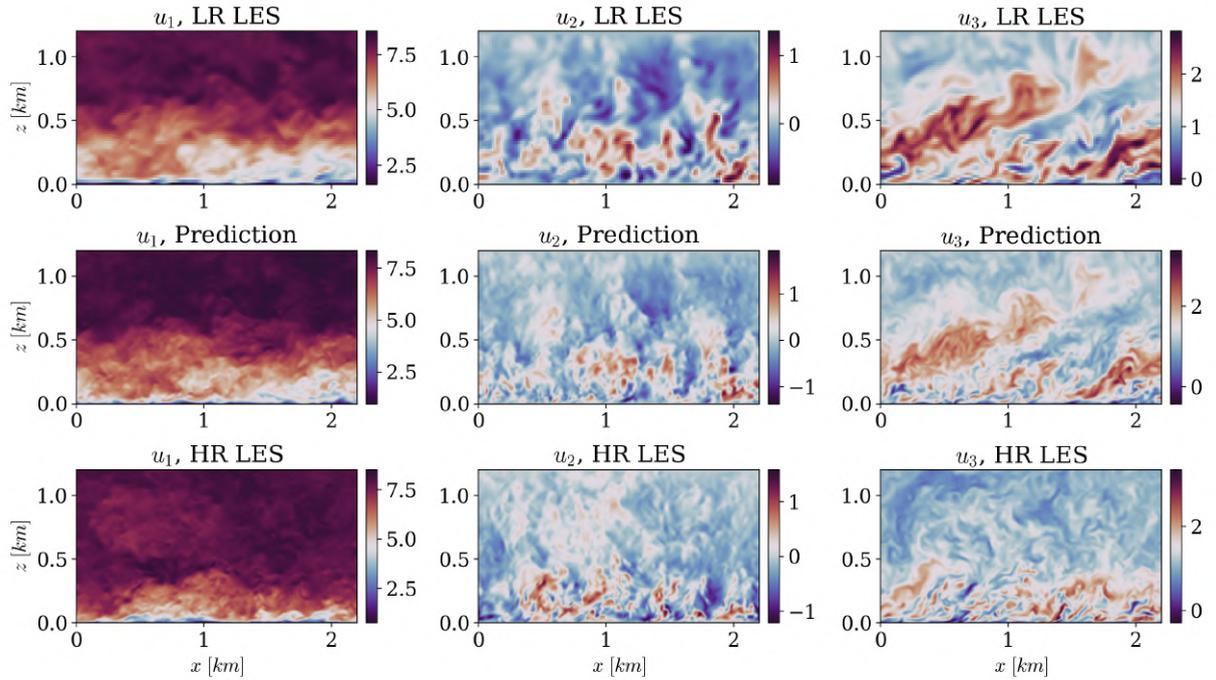

(a) Velocity components contours of low resolution LES, model prediction, and high resolution LES.

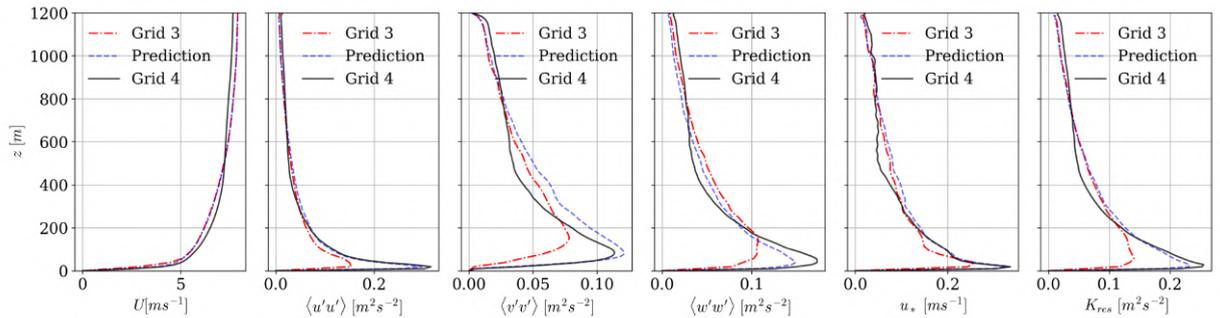

(b) Velocity profile and flow stresses of the low resolution LES, model prediction, and high resolution LES.

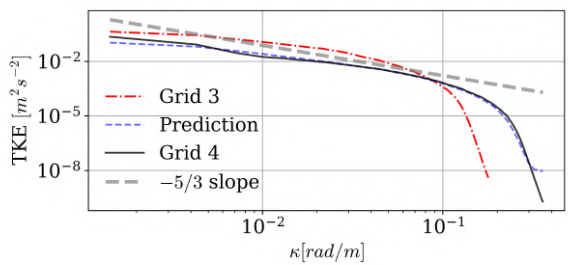

(c) Energy spectrum at altitude $z = 100$ m

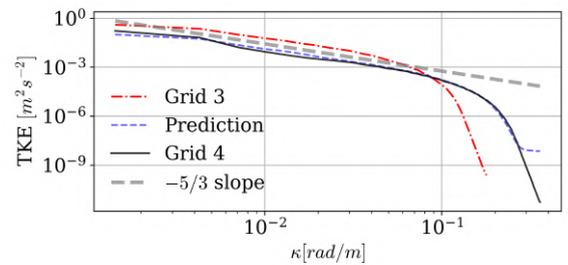

(d) Energy spectrum at altitude $z = 550$ m

**Figure 7:** Results for the interpolation task, scale factor 2 with input grid 2, geostrophic wind speed 7.15 [m/sec], surface roughness coefficient 0.05 [m]

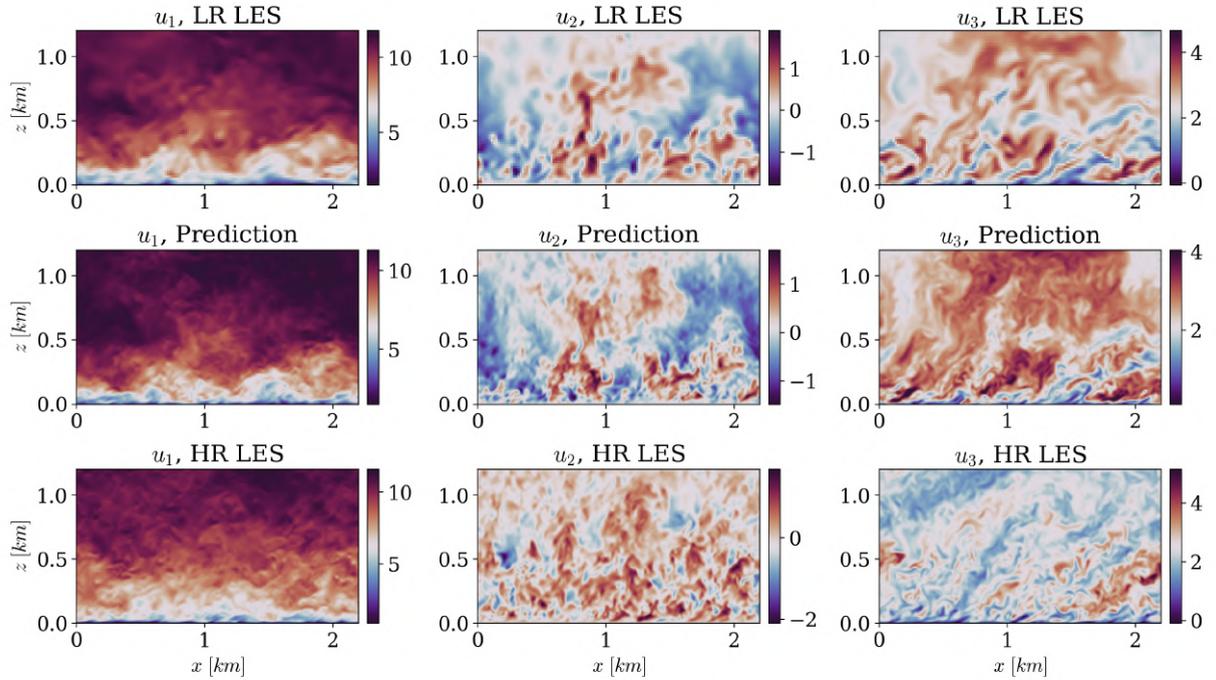

(a) Velocity components contours of low resolution LES, model prediction, and high resolution LES.

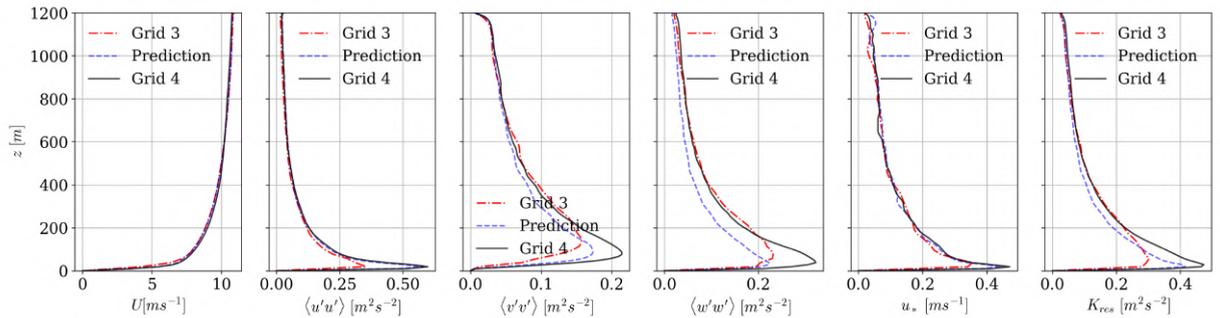

(b) Velocity profile and flow stresses of the low resolution LES, model prediction, and high resolution LES.

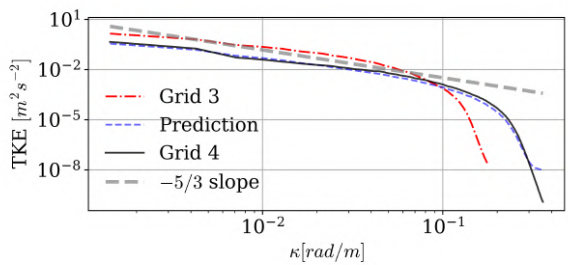 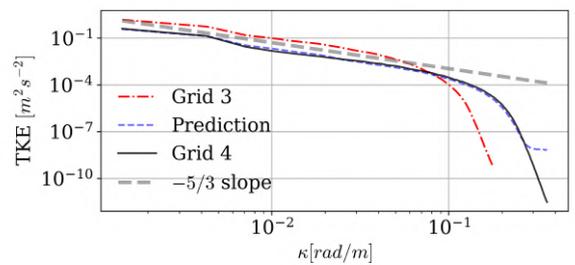

(c) Energy spectrum at altitude $z = 100$ m  (d) Energy spectrum at altitude $z = 550$ m

**Figure 8:** Results for the interpolation task, scale factor 2 with input grid 2, geostrophic wind speed 10 [m/sec], surface roughness coefficient 0.05 [m]

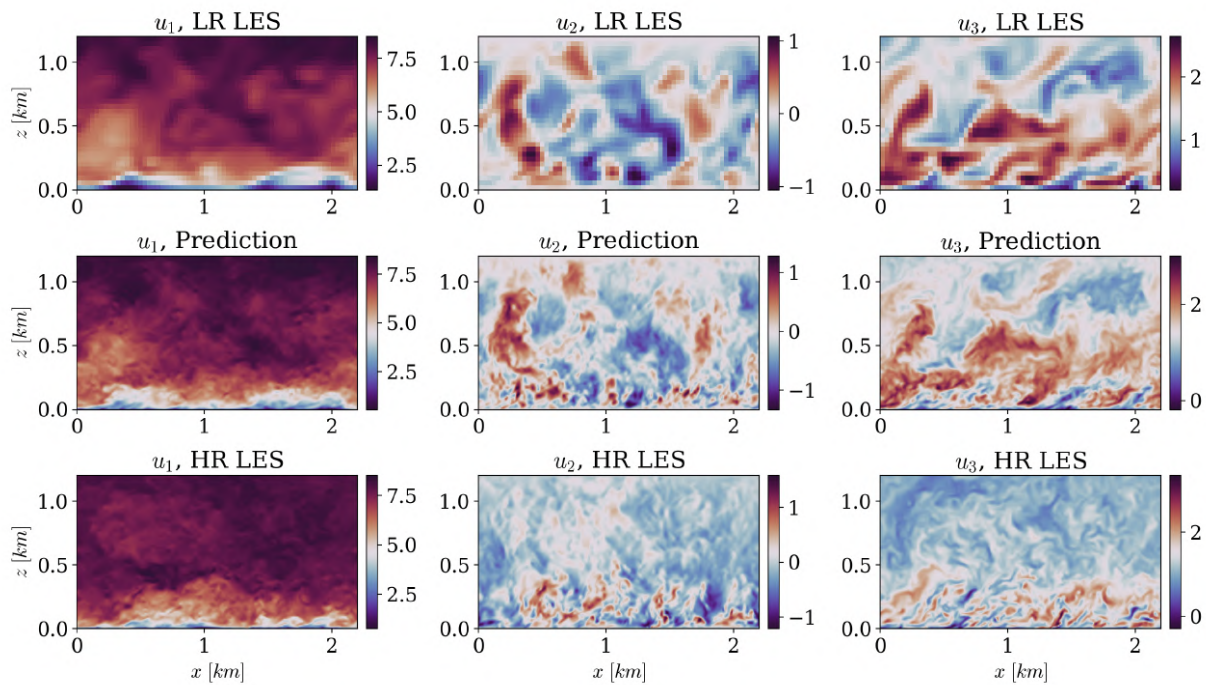

(a) Velocity components contours of low resolution LES, model prediction, and high resolution LES.

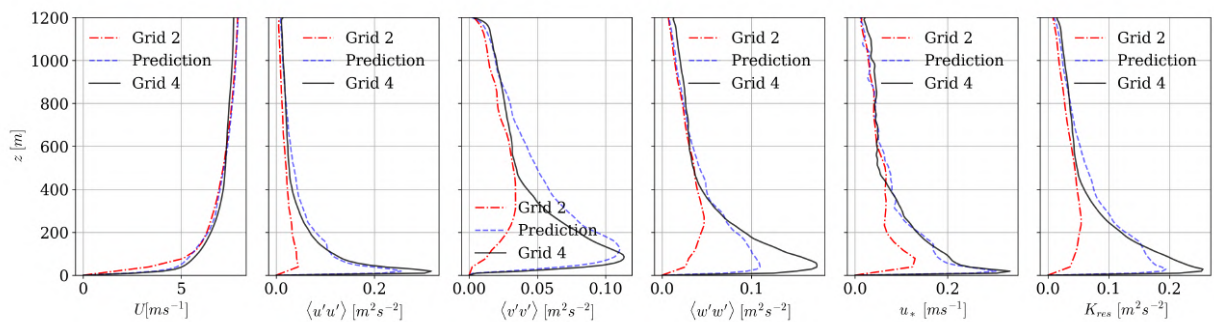

(b) Velocity profile and flow stresses of the low resolution LES, model prediction, and high resolution LES.

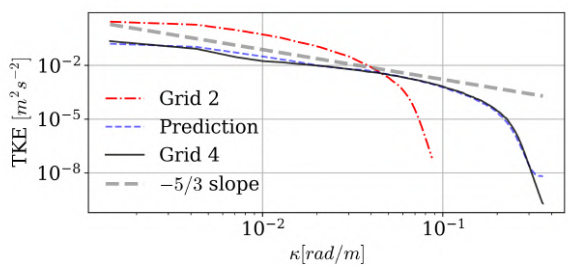

(c) Energy spectrum at altitude $z = 100$ m

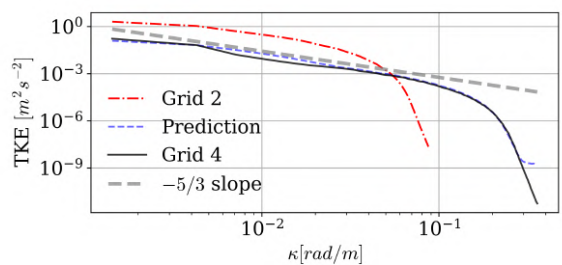

(d) Energy spectrum at altitude $z = 550$ m

**Figure 9:** Results for the interpolation task, scale factor 4 with input grid 1, geostrophic wind speed 7.15 [m/sec], surface roughness coefficient 0.05 [m]

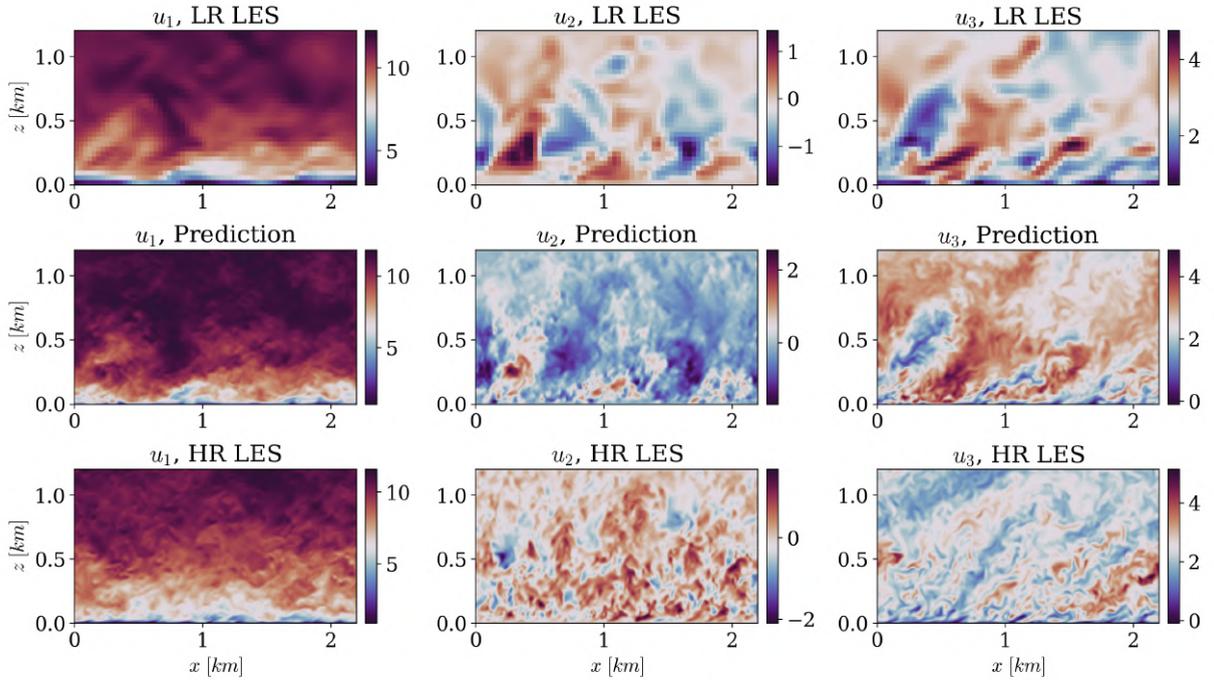

(a) Velocity components contours of low resolution LES, model prediction, and high resolution LES.

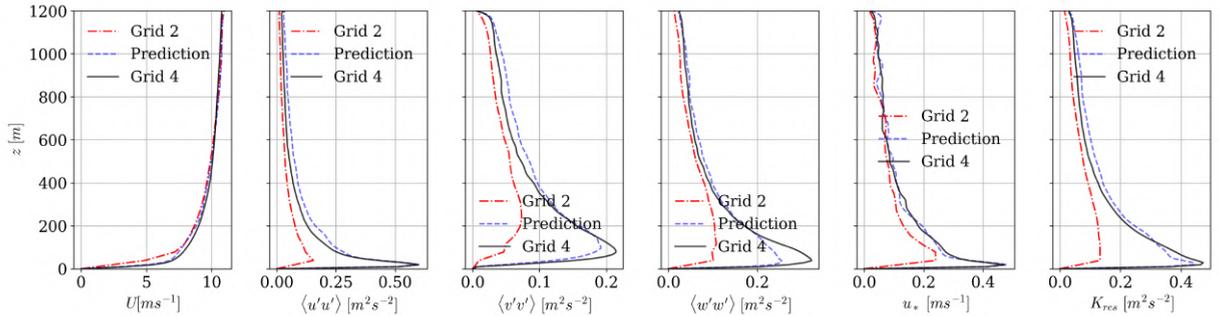

(b) Velocity profile and flow stresses of the low resolution LES, model prediction, and high resolution LES.

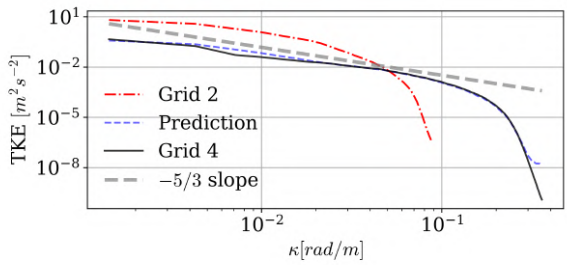

(c) Energy spectrum at altitude $z = 100$ m

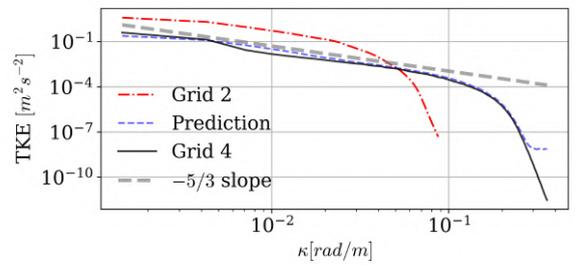

(d) Energy spectrum at altitude $z = 550$ m

**Figure 10:** Results for the interpolation task, scale factor 4 with input grid 1, geostrophic wind speed 10 [m/sec], surface roughness coefficient 0.05 [m]

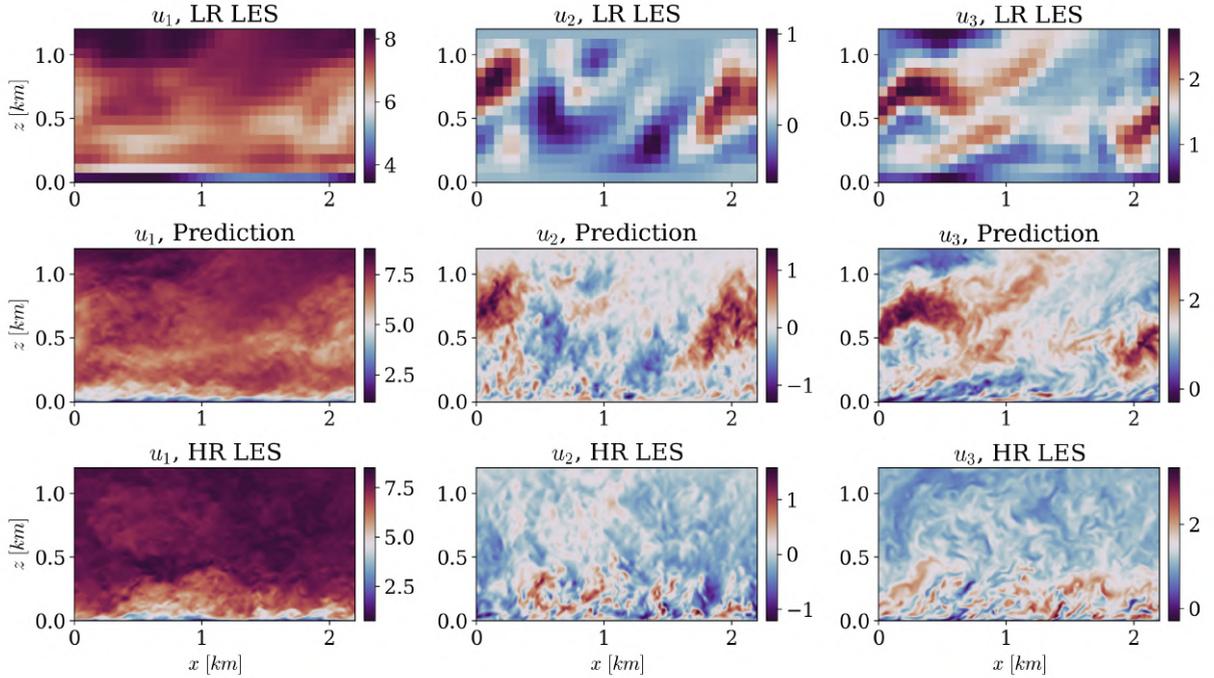

(a) Velocity components contours of low resolution LES, model prediction, and high resolution LES.

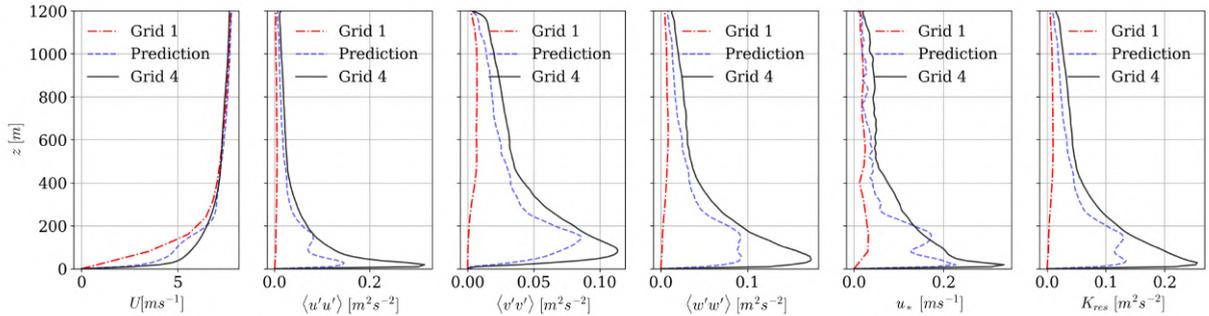

(b) Velocity profile and flow stresses of the low resolution LES, model prediction, and high resolution LES.

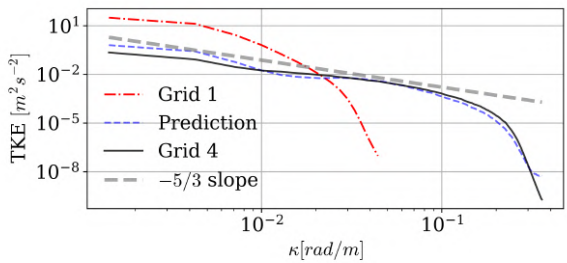

(c) Energy spectrum at altitude $z = 100$ m

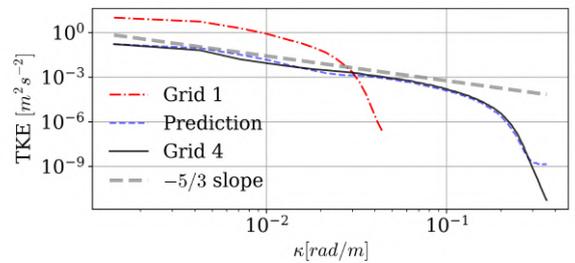

(d) Energy spectrum at altitude $z = 550$ m

**Figure 11:** Results for the interpolation task, scale factor 8 with input grid 0, geostrophic wind speed 7.15 [m/sec], surface roughness coefficient 0.05 [m]

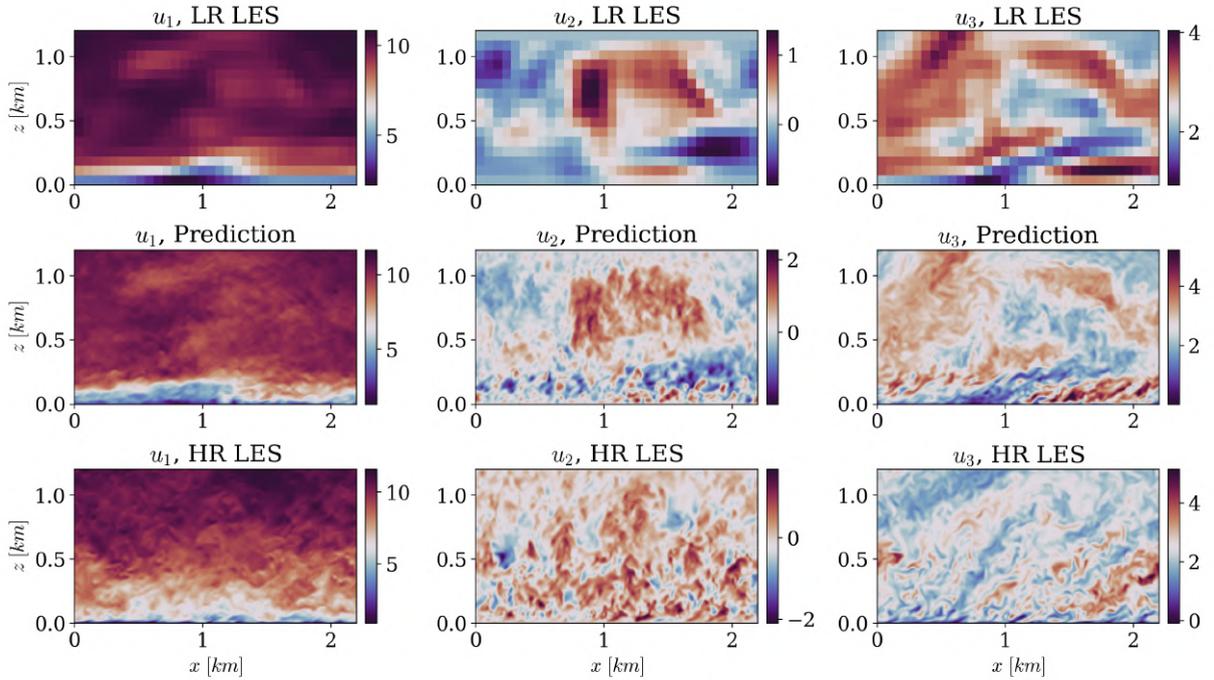

(a) Velocity components contours of low resolution LES, model prediction, and high resolution LES.

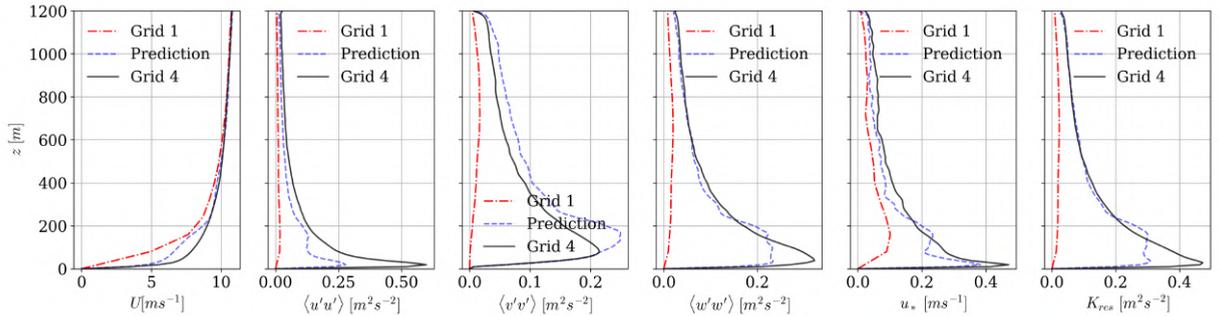

(b) Velocity profile and flow stresses of the low resolution LES, model prediction, and high resolution LES.

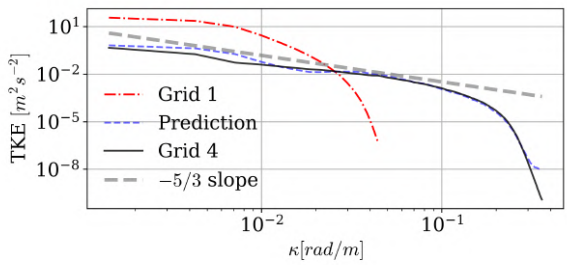

(c) Energy spectrum at altitude $z = 100$ m

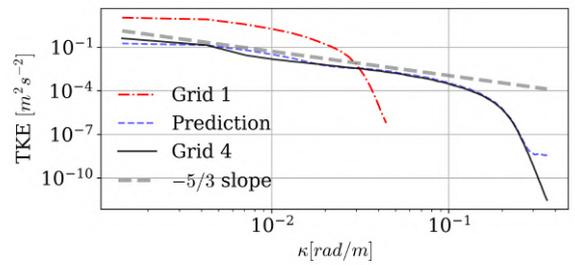

(d) Energy spectrum at altitude $z = 550$ m

**Figure 12:** Results for the interpolation task, scale factor 8 with input grid 0, geostrophic wind speed 10 [m/sec], surface roughness coefficient 0.05 [m]

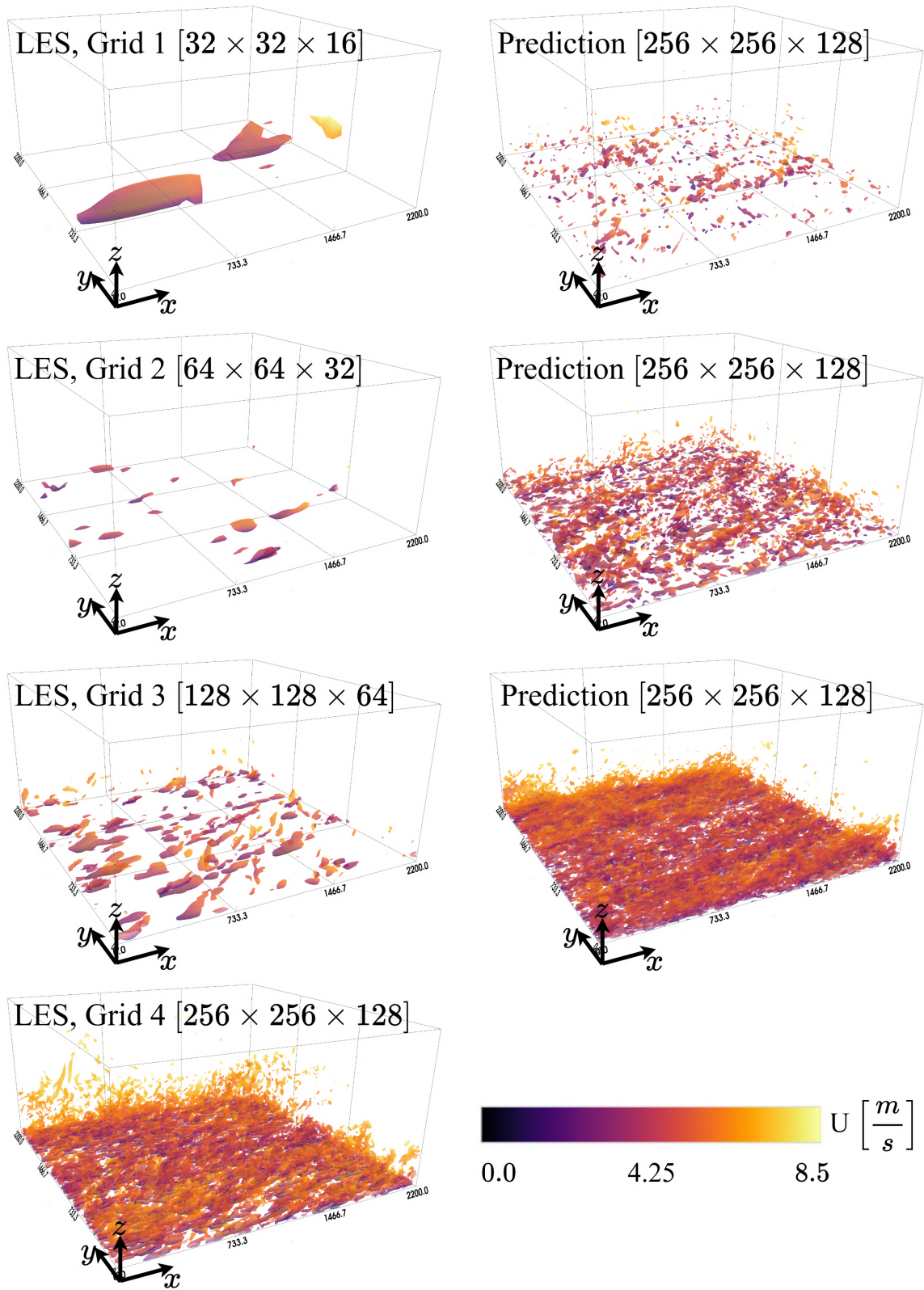

**Figure 13:** Isosurfaces of the *Q*-criterion at different input grid resolution at geostrophic wind speed 7.15 [m/sec] [Interpolation task].

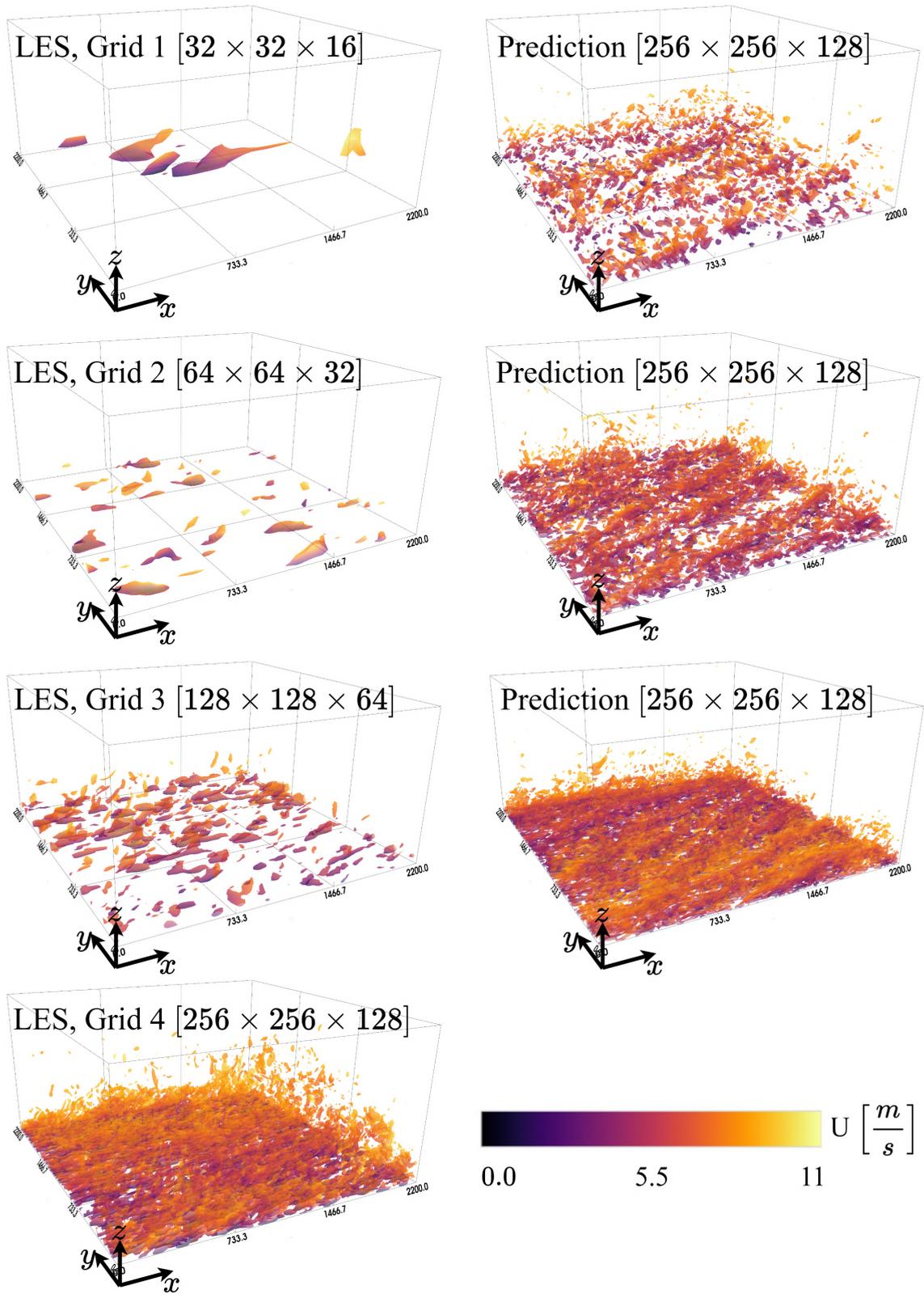

**Figure 14:** Isosurfaces of the $Q$-criterion at different input grid resolution at geostrophic wind speed 10.0 [m/sec] [Interpolation task].

## 4.2. Extrapolation task

In this section, the performance of the proposed super-resolution framework is evaluated under the extrapolation task summarized in Table 5. The model is tested on flow regimes that lie outside the parameter matrix used during training, providing a more stringent assessment of its generalization capability.

The results are presented for geostrophic wind speeds $U_g = 7.15$ m/s and $U_g = 10.0$ m/s across scale factors ×2, ×4, and ×8. These scale factors correspond to input grids 2, 1, and 0, respectively, as summarized in Table 3. This setup allows systematic evaluation of the model's robustness under increasing resolution gaps while simultaneously operating in an extrapolative parameter regime.

For the extrapolation task, the model behavior differs noticeably from the interpolation setting. As expected, the reconstruction quality deteriorates progressively as the scale factor increases. In general, the model demonstrates stronger performance at scale factor ×2 compared to ×4 and ×8. This trend is particularly evident for the moderate geostrophic wind speed $U_g = 7.15$ m/s, which lies within the wind-speed range encountered during training, albeit under different surface roughness coefficients.

In contrast, for the higher geostrophic wind speed $U_g = 10.0$ m/s, which was excluded from the training dataset, the model exhibits nonphysical behavior characterized by overestimation or underestimation of turbulent stresses and amplification of vortical structures. This degradation indicates reduced robustness when simultaneously extrapolating in both flow intensity and resolution.

For instance, Figure 15 illustrates that at scale factor ×2 and $U_g = 7.15$ m/s, the model achieves improved reconstruction of turbulent stress components, resolved turbulent kinetic energy, and energy spectra compared to the low-resolution input. The velocity field visualization further confirms that coherent small-scale structures are restored while maintaining physically consistent boundary-layer organization.

On the other hand, Figure 16 shows that for scale factor ×2 and $U_g = 10.0$ m/s, the model overestimates vertical turbulent stress magnitudes and underestimate the lateral stresses. The corresponding energy spectrum exhibits amplified high-wavenumber content, and the reconstructed velocity field contains noisy small-scale fluctuations with reduced structural coherence. This behavior suggests that the model amplifies turbulence intensity when operating outside its trained wind-speed regime.

For scale factor ×4 at the moderate geostrophic wind speed $U_g = 7.15$ m/s, the model demonstrates clear improvement in the reconstruction of turbulent stresses and energy spectra relative to the low-resolution input. As shown in Figure 17, the predicted flow patterns remain comparable to the high-resolution LES reference, and the overall boundary-layer structure is preserved with physically consistent shear distribution.

However, at the higher wind speed $U_g = 10.0$ m/s, the model exhibits systematic underestimation of the turbulent stress components, as illustrated in Figure 18. This behavior contrasts with the scale factor ×2 case, where stress overestimation was observed. The underprediction at ×4 suggests that, under combined resolution and parameter extrapolation, the model adopts a more conservative reconstruction strategy, leading to damped fluctuation amplitudes.

A similar trend is observed for scale factor ×8. The reconstructed turbulent stresses are consistently underestimated for both $U_g = 7.15$ m/s and $U_g = 10.0$ m/s. The increased resolution gap limits the recovery of high-frequency turbulent content, resulting in reduced stress magnitudes and attenuated spectral energy across the inertial subrange.

Despite the overestimation or underestimation observed in the lateral and vertical Reynolds stress components, the streamwise stress component remains the most accurately reconstructed across all scale factors. This component is particularly significant for wind turbine load assessment, as it directly influences fatigue loading and structural response according to IEC standards, as discussed in [4]. The relatively robust reconstruction of the streamwise stress therefore indicates that the model preserves the most critical turbulence metric for wind energy applications, even under extrapolation conditions.

Figures 21 and 22 present the isosurfaces of the $Q$-criterion for the low-resolution LES input, the model reconstruction, and the high-resolution reference across different input grid resolutions for geostrophic wind speeds $U_g = 7.15$ m/s and $U_g = 10.0$ m/s.

For the intermediate geostrophic wind speed ($U_g = 7.15$ m/s, Figure 21), the model exhibits behavior consistent with the interpolation regime. At lower input resolutions, the reconstructed isosurfaces show reduced structural density and weaker vortex coherence, reflecting the limited turbulent information available in the coarse input field. As the input grid resolution increases (corresponding to smaller scale factors), the reconstructed $Q$-criterion structures become progressively denser and more spatially organized, approaching the topology and distribution observed in the high-resolution LES.

In contrast, for the higher geostrophic wind speed ($U_g$ = 10.0 m/s, Figure 22), which lies further outside the training distribution, the predicted isosurfaces at input grid 3 display noisy and overestimated velocity turbulent structures compared to the LES reference, indicating reduced generalization capability under stronger extrapolation conditions.

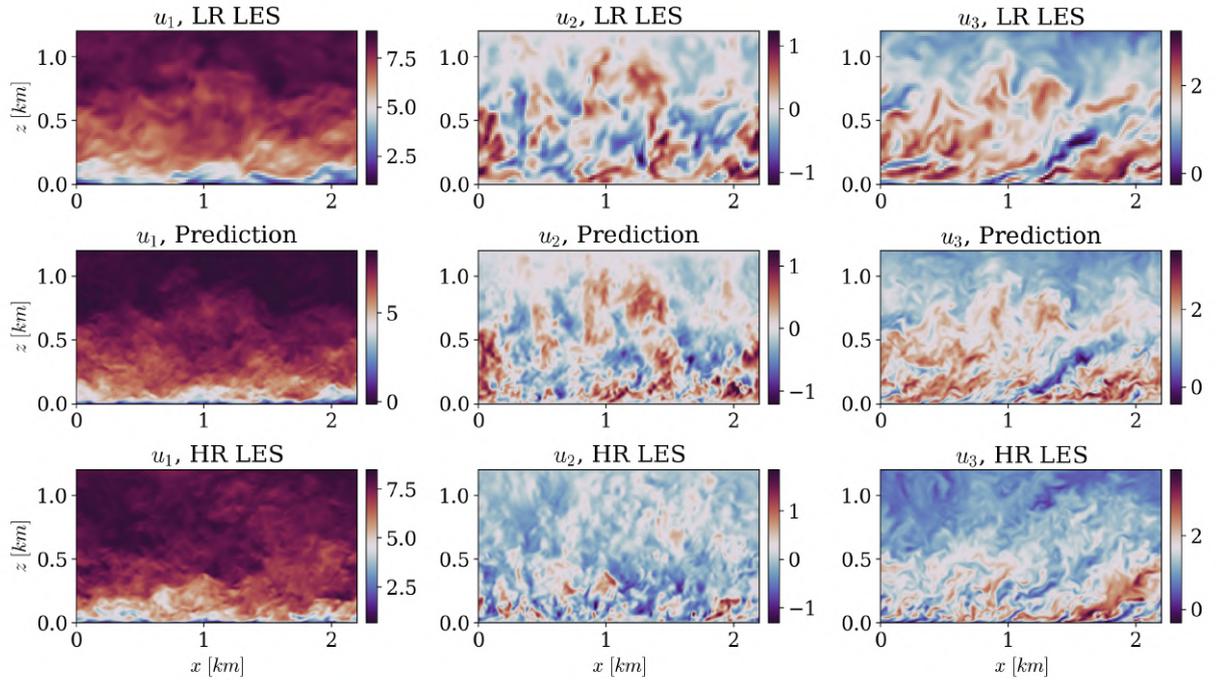

(a) Velocity components contours of low resolution LES, model prediction, and high resolution LES.

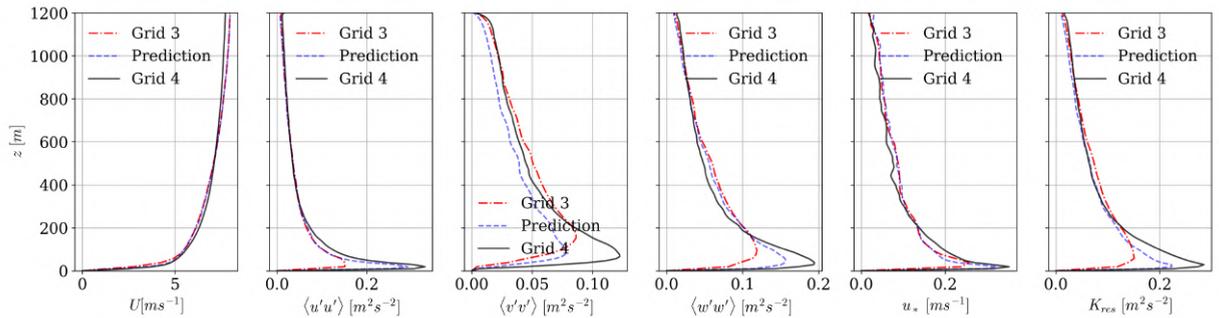

(b) Velocity profile and flow stresses of the low resolution LES, model prediction, and high resolution LES.

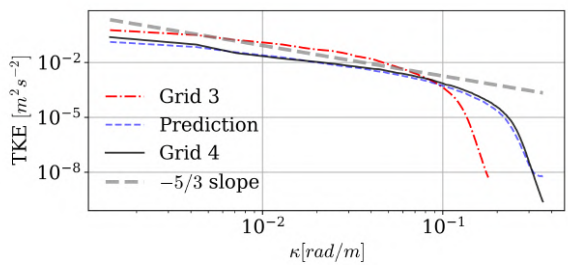

(c) Energy spectrum at altitude $z = 100$ m

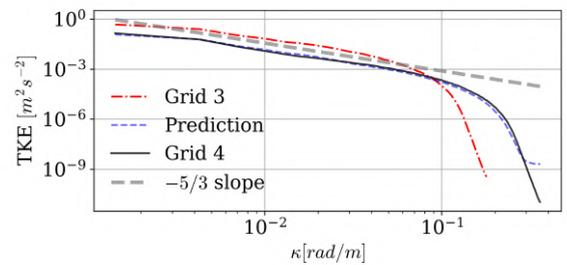

(d) Energy spectrum at altitude $z = 550$ m

**Figure 15:** Results for the extrapolation task, scale factor 2 with input grid 2, geostrophic wind speed 7.15 [m/sec], surface roughness coefficient 0.1 [m]

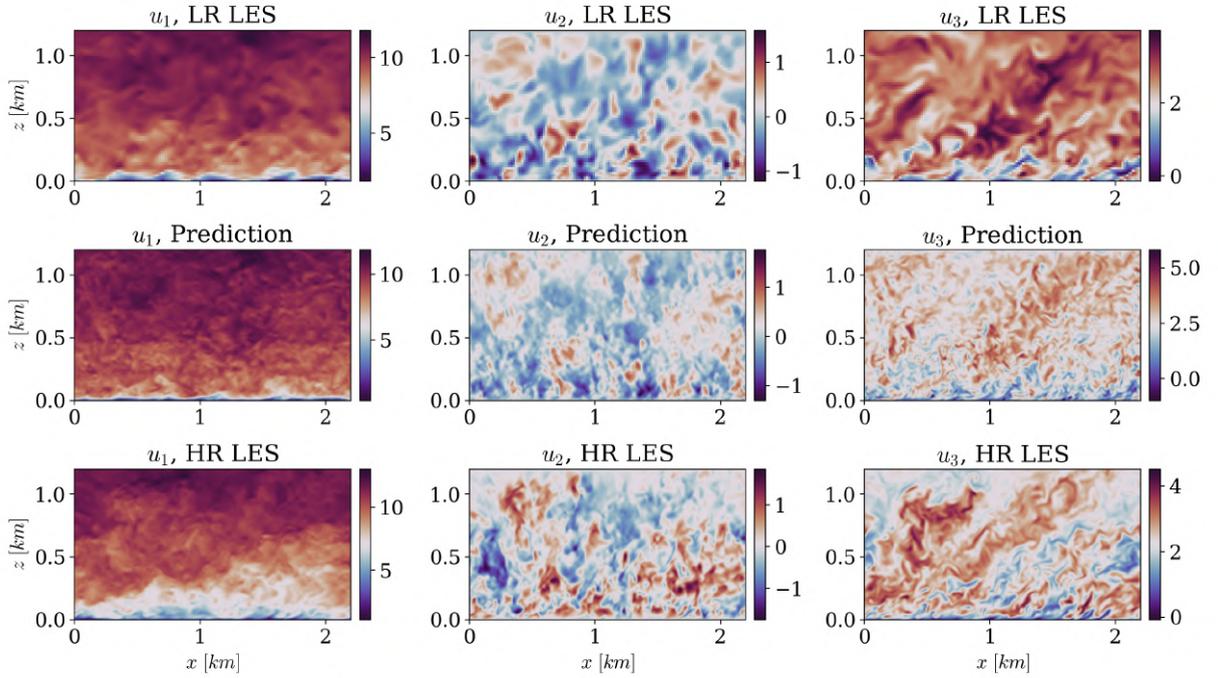

(a) Velocity components contours of low resolution LES, model prediction, and high resolution LES.

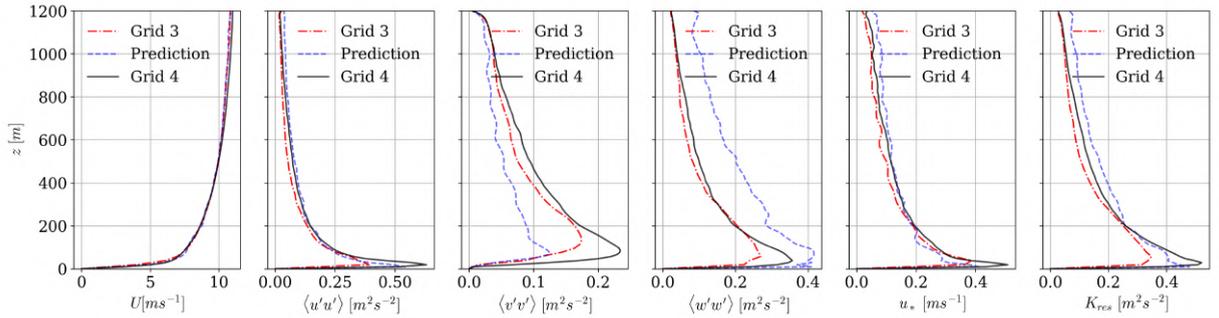

(b) Velocity profile and flow stresses of the low resolution LES, model prediction, and high resolution LES.

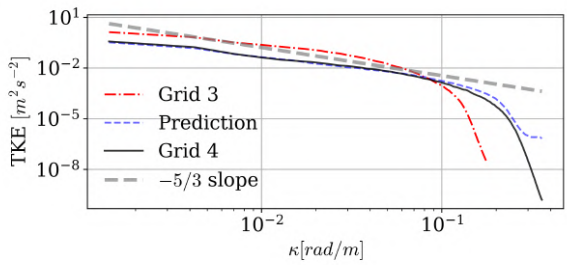

(c) Energy spectrum at altitude $z = 100$ m

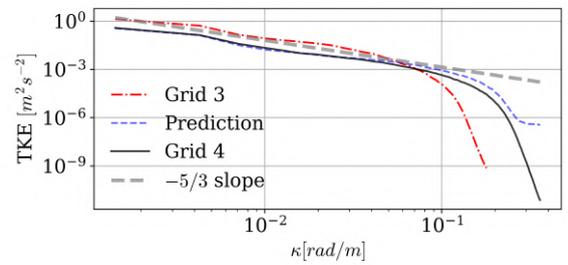

(d) Energy spectrum at altitude $z = 550$ m

**Figure 16:** Results for the extrapolation task, scale factor 2 with input grid 2, geostrophic wind speed 10 [m/sec], surface roughness coefficient 0.1 [m]

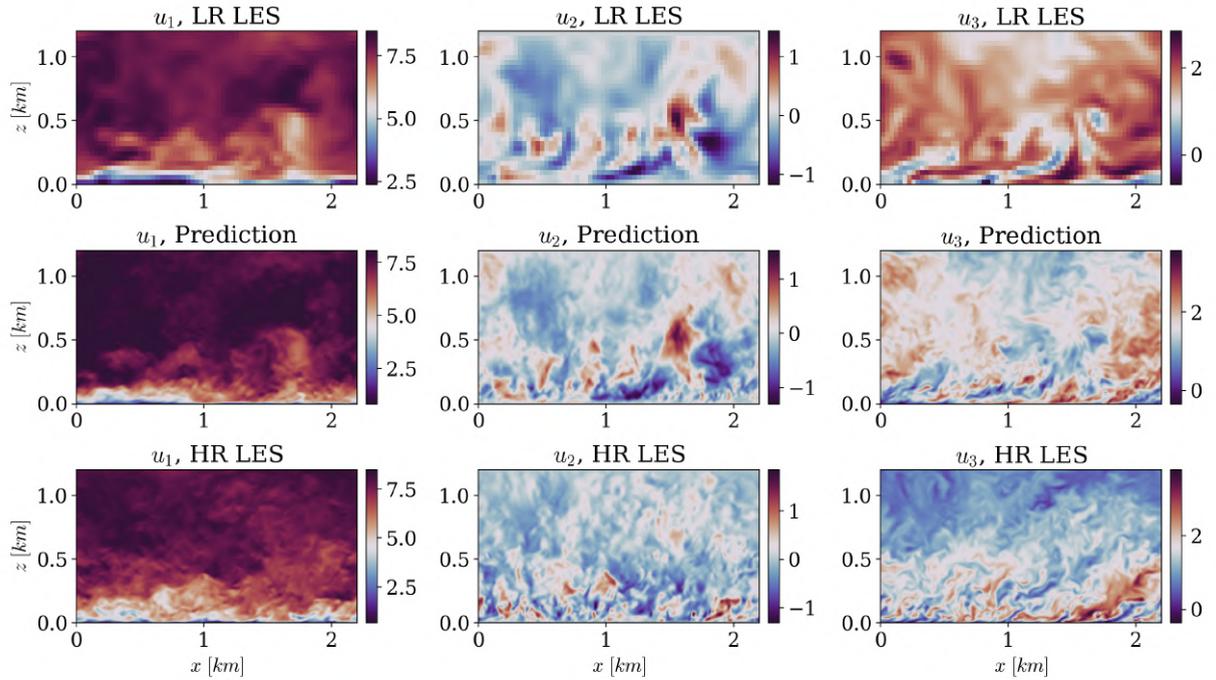

(a) Velocity components contours of low resolution LES, model prediction, and high resolution LES.

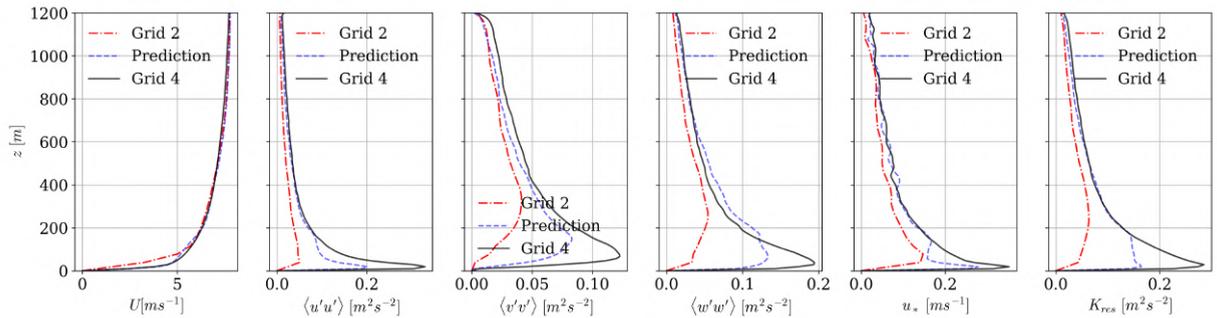

(b) Velocity profile and flow stresses of the low resolution LES, model prediction, and high resolution LES.

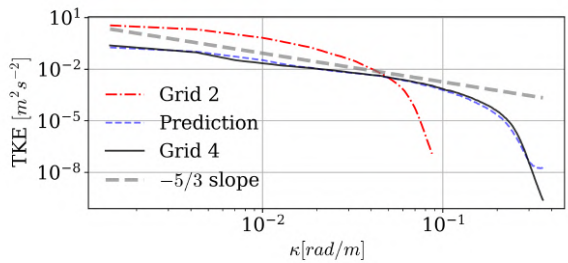

(c) Energy spectrum at altitude $z = 100$ m

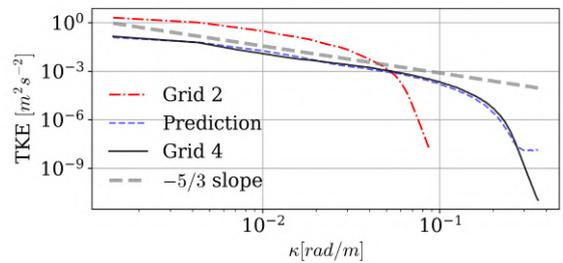

(d) Energy spectrum at altitude $z = 550$ m

**Figure 17:** Results for the extrapolation task, scale factor 4 with input grid 1, geostrophic wind speed 7.15 [m/sec], surface roughness coefficient 0.1 [m]

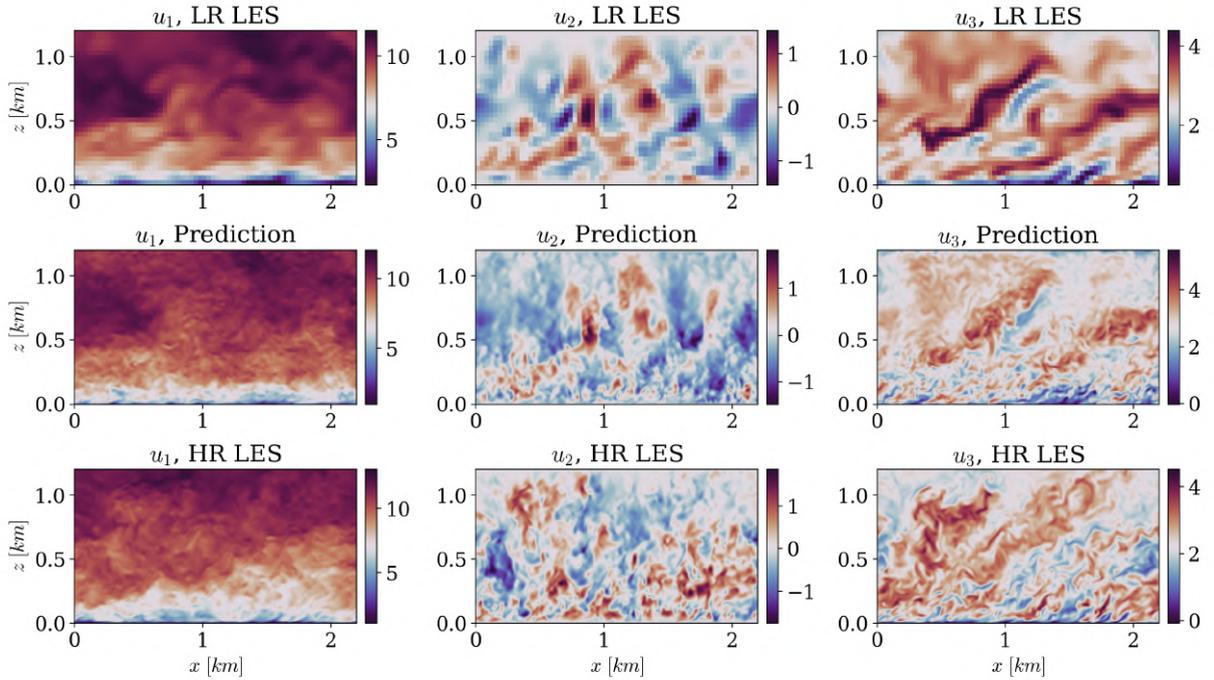

(a) Velocity components contours of low resolution LES, model prediction, and high resolution LES.

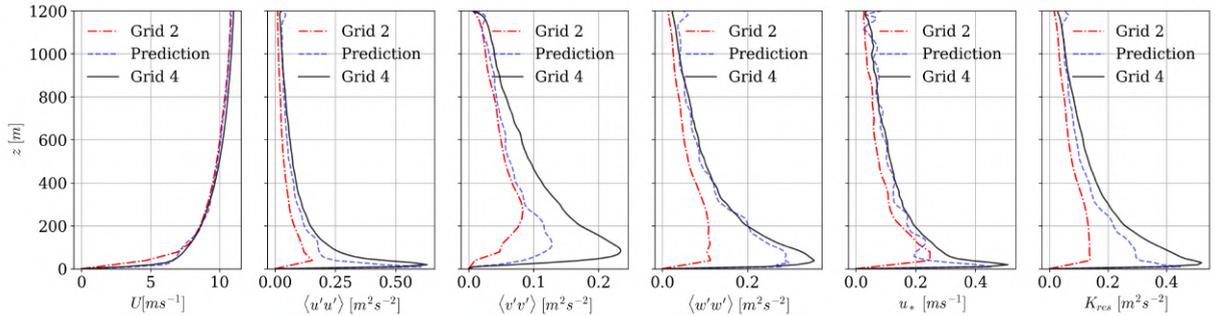

(b) Velocity profile and flow stresses of the low resolution LES, model prediction, and high resolution LES.

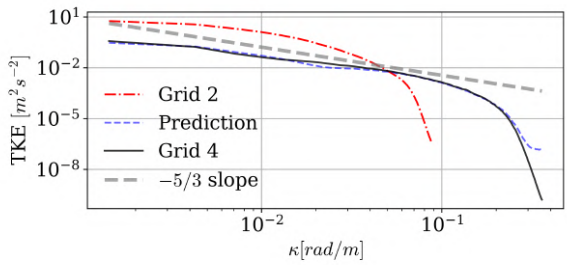

(c) Energy spectrum at altitude $z = 100$ m

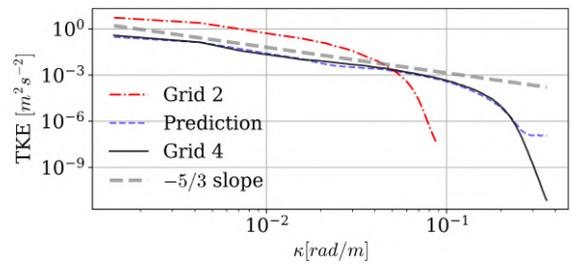

(d) Energy spectrum at altitude $z = 550$ m

**Figure 18:** Results for the extrapolation task, scale factor 4 with input grid 1, geostrophic wind speed 10 [m/sec], surface roughness coefficient 0.1 [m]

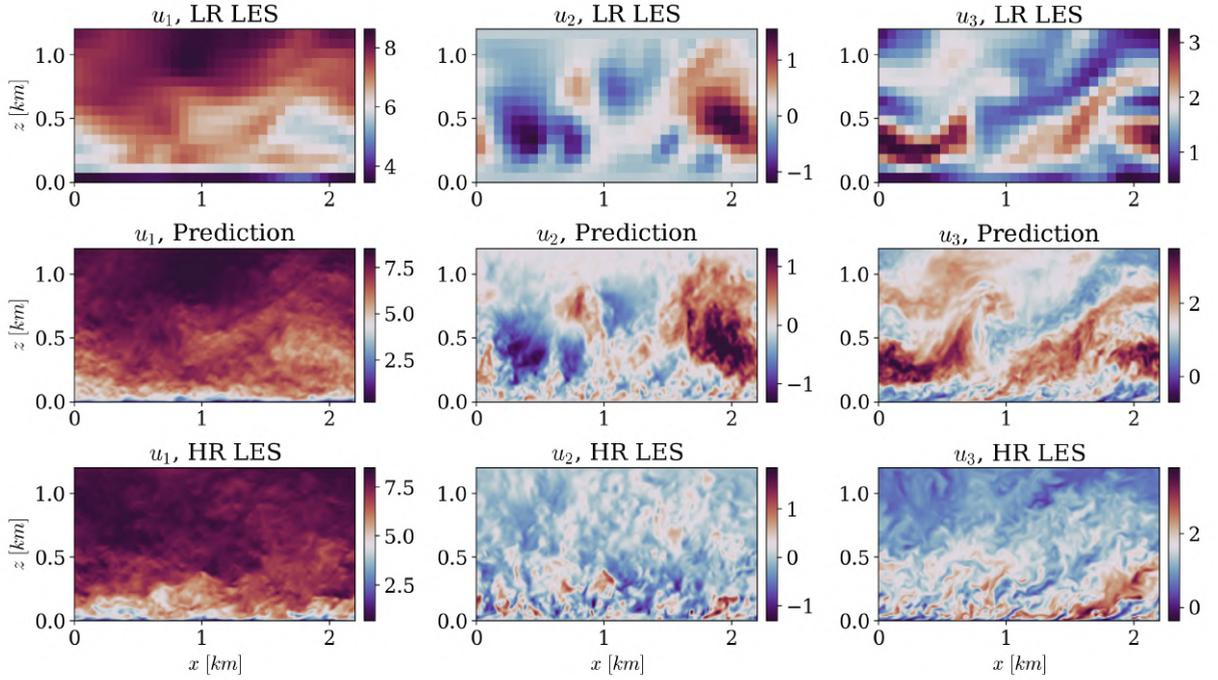

(a) Velocity components contours of low resolution LES, model prediction, and high resolution LES.

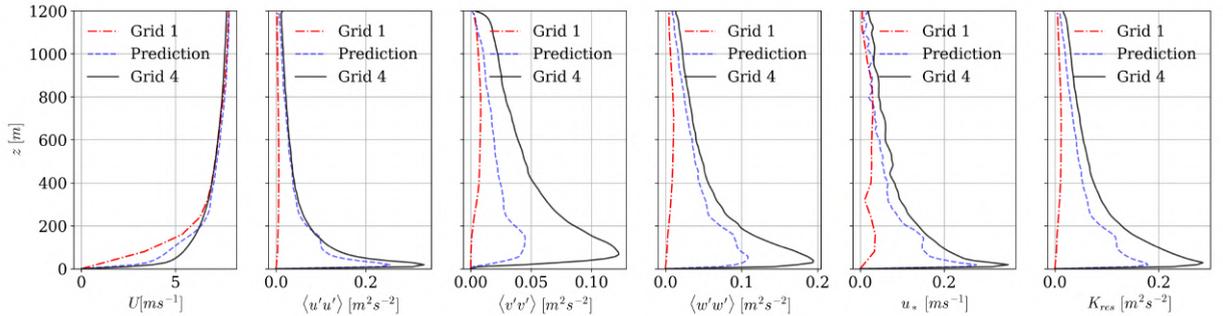

(b) Velocity profile and flow stresses of the low resolution LES, model prediction, and high resolution LES.

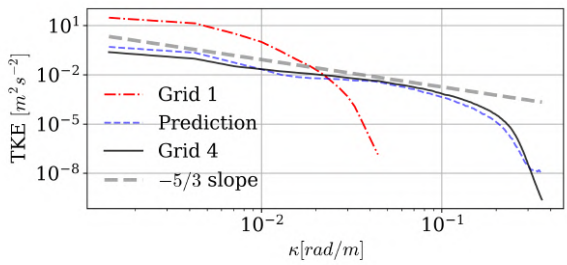

(c) Energy spectrum at altitude $z = 100$ m

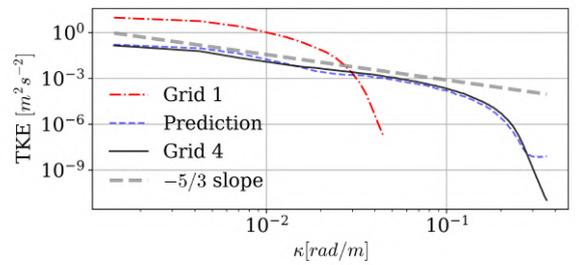

(d) Energy spectrum at altitude $z = 550$ m

**Figure 19:** Results for the extrapolation task, scale factor 8 with input grid 0, geostrophic wind speed 7.15 [m/sec], surface roughness coefficient 0.1 [m]

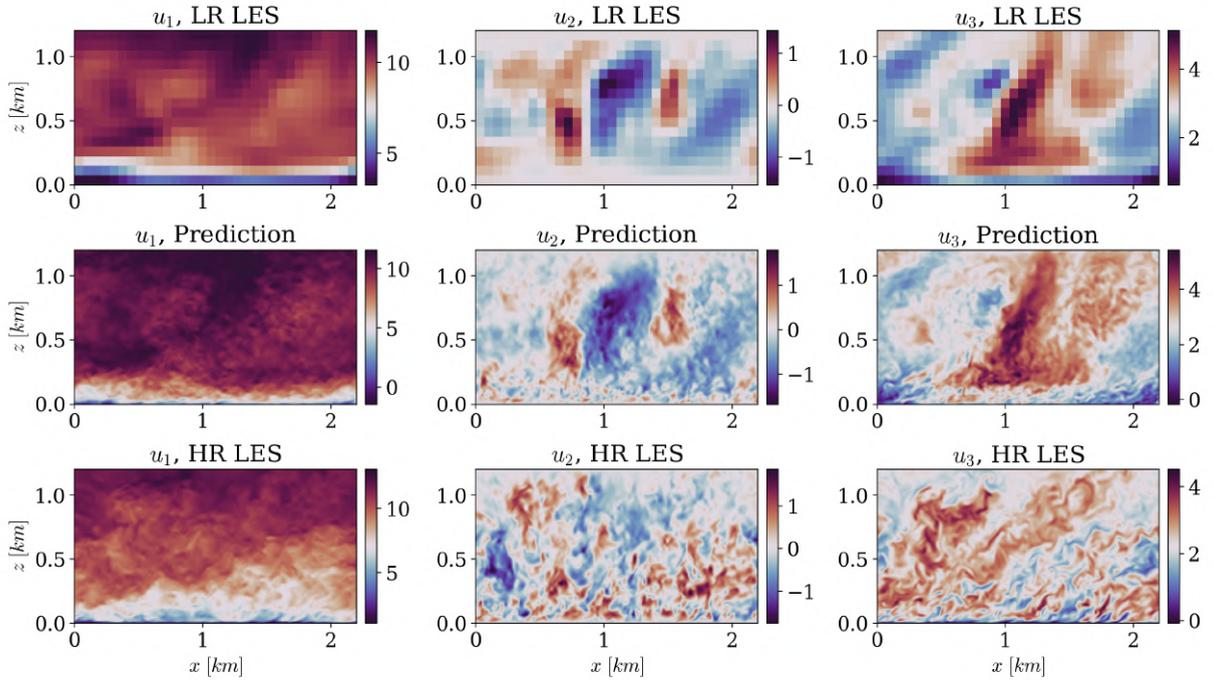

(a) Velocity components contours of low resolution LES, model prediction, and high resolution LES.

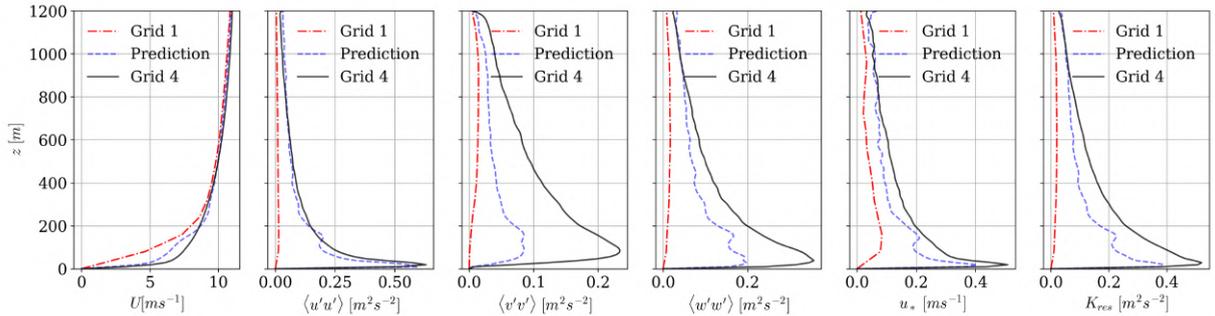

(b) Velocity profile and flow stresses of the low resolution LES, model prediction, and high resolution LES.

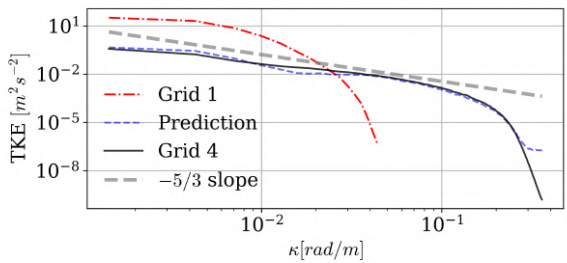

(c) Energy spectrum at altitude $z = 100$ m

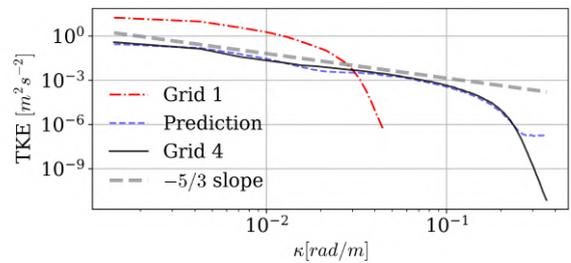

(d) Energy spectrum at altitude $z = 550$ m

**Figure 20:** Results for the extrapolation task, scale factor 8 with input grid 0, geostrophic wind speed 10 [m/sec], surface roughness coefficient 0.1 [m]

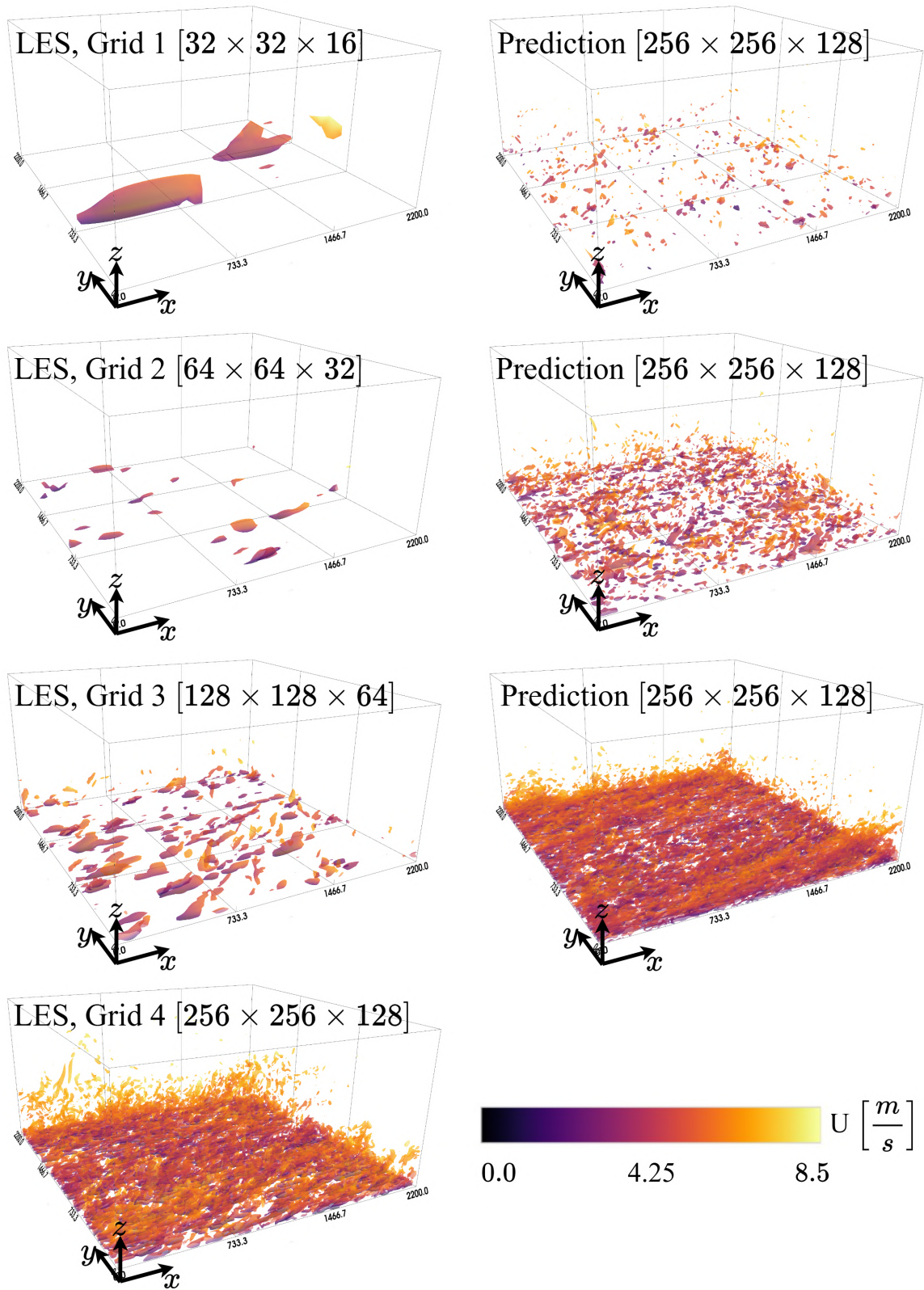

**Figure 21:** Isosurfaces of the $Q$-criterion at different input grid resolution at geostrophic wind speed 7.15 [m/sec] [Extrapolation task].

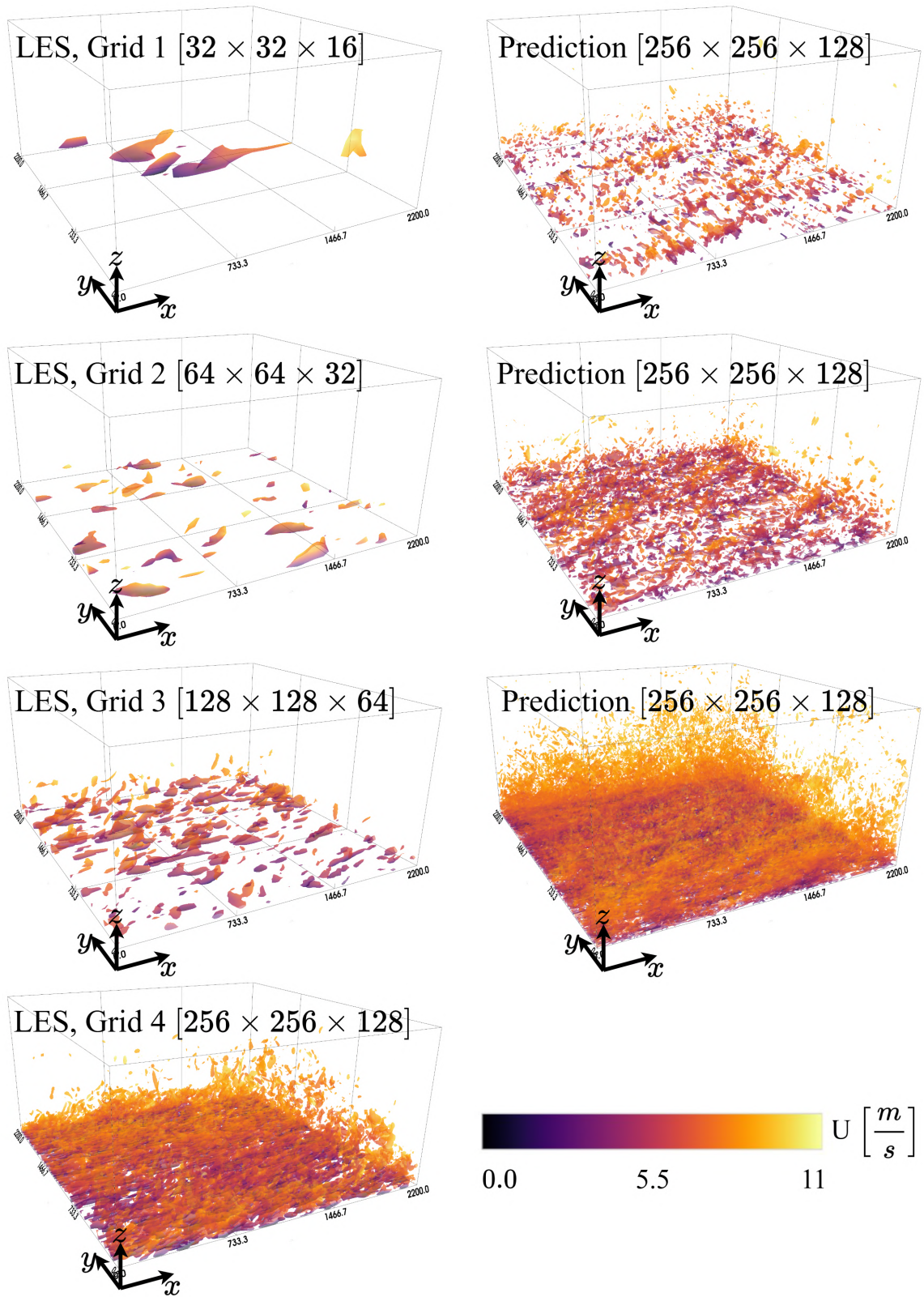

**Figure 22:** Isosurfaces of the $Q$-criterion at different input grid resolution at geostrophic wind speed 10.0 [m/sec] [Extrapolation task].

# 5. Conclusion

sec:conclusion

This study presented a conditional denoising diffusion probabilistic model for the super-resolution of atmospheric boundary layer (ABL) large eddy simulations (LES). A high-fidelity numerical dataset was generated using the `Xcompact3D` solver, based on a high-order finite-difference framework, spanning a range of geostrophic wind speeds and surface roughness conditions aligned with IEC wind classes, and across multiple grid resolutions. This dataset enabled a systematic assessment of the model under controlled physical variability.

The proposed diffusion framework was trained to reconstruct fine-scale turbulent structures from coarse-resolution inputs for different super-resolution scale factors. Its performance was evaluated under both interpolation and extrapolation scenarios. For interpolation tasks, the model demonstrated strong capability in recovering key turbulent characteristics, including coherent flow structures and Reynolds stresses, indicating that the learned mappings preserve essential physical features within the training regime.

In contrast, extrapolation to higher geostrophic wind speeds revealed notable limitations, characterized by increased noise levels and overestimation of turbulent stresses. These findings highlight the challenges associated with generalization beyond the range of training conditions, particularly for highly nonlinear turbulent flows.

Overall, this work demonstrates the potential of physics-informed generative models for turbulent flow super-resolution. The proposed approach offers a computationally efficient tool that significantly reduce the inference time while retaining physically consistent predictions within the training domain. This capability is particularly relevant for wind energy and environmental engineering applications, where rapid and accurate characterization of turbulent flow fields is essential.

**Data availability**

The code and the data required to reproduce the presented results will be publicly available on GitHub after the peer revision is complete.

**Declaration of competing interest**

The authors declare that they have no known competing financial interests or personal relationships that could have appeared to influence the work reported in this paper.